\documentclass[floatfix]{iopart}
\usepackage{graphicx}
\pdfoutput=1
\usepackage{epstopdf}
\usepackage{bm}
\usepackage{amssymb}

\begin{document}
\newcommand\Hbar{$\bar{\rm H}$}
\newcommand\Hbars{$\bar{\rm H}$s}
\newcommand\pbar{$\bar{\rm p}$}
\newcommand\pbars{$\bar{\rm p}\,$'s}
\newcommand\pos{e$^+$}
\newcommand\poss{e$^+\,$'s}
\newcommand\elec{e$^-$}

\newcommand\kB{k_\mathrm{B}}
\newcommand\Rw{R_\mathrm{W}}
\newcommand\Energy{\mathcal E}
\newcommand\Eperp{\Energy_\perp}
\newcommand\Eperpzero{\Energy_{\perp 0}}
\newcommand\EperpTrap{\Energy_{\perp\mathrm{MirTrap}}}
\newcommand\ETrap{\Energy_{\mathrm{Trap}}}
\newcommand\Eperpone{\Energy_{\perp 1}}
\newcommand\Epar{\Energy_{\parallel}}
\newcommand\Eparzero{\Energy_{\parallel 0}}
\newcommand\Eparone{\Energy_{\parallel 1}}
\newcommand\Emax{E_\mathrm{max}}
\newcommand\Ebind{\Energy_\mathrm{b}}
\newcommand\Ekin{\Energy_\mathrm{kin}}
\newcommand\muHbar{\mu_{\bar \mathrm{H}}}
\newcommand\muHbarb{\bm{\mu}_{\bar \mathrm{H}}}
\newcommand\mupbar{\mu_{\bar \mathrm{p}}}
\newcommand\mupbarb{\bm{\mu}_{\bar \mathrm{p}}}
\newcommand\Rpbar{R_{\bar \mathrm{p}}}
\newcommand\Rpbarc{{\mathcal R}_{{\bar \mathrm{p}}\mathrm{c}}}
\newcommand\Bwall{B_{\mathrm{wall}}}
\newcommand\Bmax{B_{\mathrm{max}}}
\newcommand\eV{\mathrm{eV}}
\newcommand\Nauto{N_\mathrm{a}}

\newcommand\rhat{{\hat \mathbf{r}}}
\newcommand\thetahat{{\hat{\bm\theta}}}
\newcommand\zhat{{\hat\mathbf{z}}}
\newcommand\xhat{{\hat\mathbf{x}}}
\newcommand\yhat{{\hat\mathbf{y}}}

\title[Antihydrogen and mirror-trapped antiproton discrimination]{Discriminating between antihydrogen and mirror-trapped antiprotons in a minimum-B trap}

\author{C Amole$^1$, G B Andresen$^2$, M D Ashkezari$^3$, M Baquero-Ruiz$^4$, W Bertsche$^5$, E Butler$^{5,6}$, C L Cesar$^7$, S Chapman$^4$, M Charlton$^5$, A Deller$^5$, S Eriksson$^5$, J Fajans$^{4,8}$, T Friesen$^9$, M C Fujiwara$^{10}$, D R Gill$^{10}$, A Gutierrez$^{11}$, J S Hangst$^2$, W N Hardy$^{11}$, M E Hayden$^3$, A J Humphries$^5$, R Hydomako$^9$, L Kurchaninov$^{10}$, S Jonsell$^{12}$, N Madsen$^5$, S Menary$^1$, P Nolan$^{13}$, K Olchanski$^{10}$, A Olin$^{10}$, A Povilus$^4$, P Pusa$^{13}$, F Robicheaux$^{14}$, E Sarid$^{15}$, D M Silveira$^7$, C So$^4$, J W Storey$^{10}$, R I Thompson$^9$, D P van der Werf$^5$, J S Wurtele$^{4, 8}$}

\address{$^1$ Department of Physics and Astronomy, York University, Toronto, ON, M3J 1P3, Canada}
\address{$^2$ Department of Physics and Astronomy, Aarhus University, DK-8000 Aarhus C, Denmark}
\address{$^3$ Department of Physics, Simon Fraser University, Burnaby BC, Canada V5A 1S6}
\address{$^4$ Department of Physics, University of California at Berkeley, Berkeley, CA 94720-7300, USA}
\address{$^5$ Department of Physics, College of Science, Swansea University, Swansea SA2 8PP, United Kingdom}
\address{$^6$ Physics Department, CERN, CH-1211 Geneva 23, Switzerland}
\address{$^7$ Instituto de F\'{i}sica, Universidade Federal do Rio de Janeiro, Rio de Janeiro 21941-972, Brazil}
\address{$^8$ Lawrence Berkeley National Laboratory, Berkeley, CA 94720, USA}
\address{$^9$ Department of Physics and Astronomy, University of Calgary, Calgary AB, Canada T2N 1N4}
\address{$^{10}$ TRIUMF, 4004 Wesbrook Mall, Vancouver BC, Canada V6T 2A3}
\address{$^{11}$ Department of Physics and Astronomy, University of British Columbia, Vancouver BC, Canada V6T 1Z4}
\address{$^{12}$ Department of Physics, Stockholm University, SE-10691, Stockholm, Sweden}
\address{$^{13}$ Department of Physics, University of Liverpool, Liverpool L69 7ZE, United Kingdom}
\address{$^{14}$ Department of Physics, Auburn University, Auburn, AL 36849-5311, USA}
\address{$^{15}$ Department of Physics, NRCN-Nuclear Research Center Negev, Beer Sheva, IL-84190, Israel}

\date{Received \today}

\begin{abstract} Recently, antihydrogen atoms were trapped at CERN in a magnetic minimum (minimum-B) trap formed by superconducting octupole and mirror magnet coils.  The trapped antiatoms were detected by rapidly turning off these magnets, thereby eliminating the magnetic minimum and releasing any antiatoms contained in the trap.  Once released, these antiatoms quickly hit the trap wall, whereupon the positrons and antiprotons in the antiatoms annihilated.  The antiproton annihilations produce easily detected signals; we used these signals to prove that we trapped antihydrogen. However, our technique could be confounded by mirror-trapped antiprotons, which would produce seemingly-identical annihilation signals upon hitting the trap wall. In this paper, we discuss possible sources of mirror-trapped antiprotons and show that antihydrogen and antiprotons can be readily distinguished, often with the aid of applied electric fields, by analyzing the annihilation locations and times.  We further discuss the general properties of antiproton and antihydrogen trajectories in this magnetic geometry, and reconstruct the antihydrogen energy distribution from the measured annihilation time history.
\end{abstract}


\maketitle

\section{Introduction}
\label{Introduction}
\begin{figure}
\centerline{\includegraphics[width=2.5in]{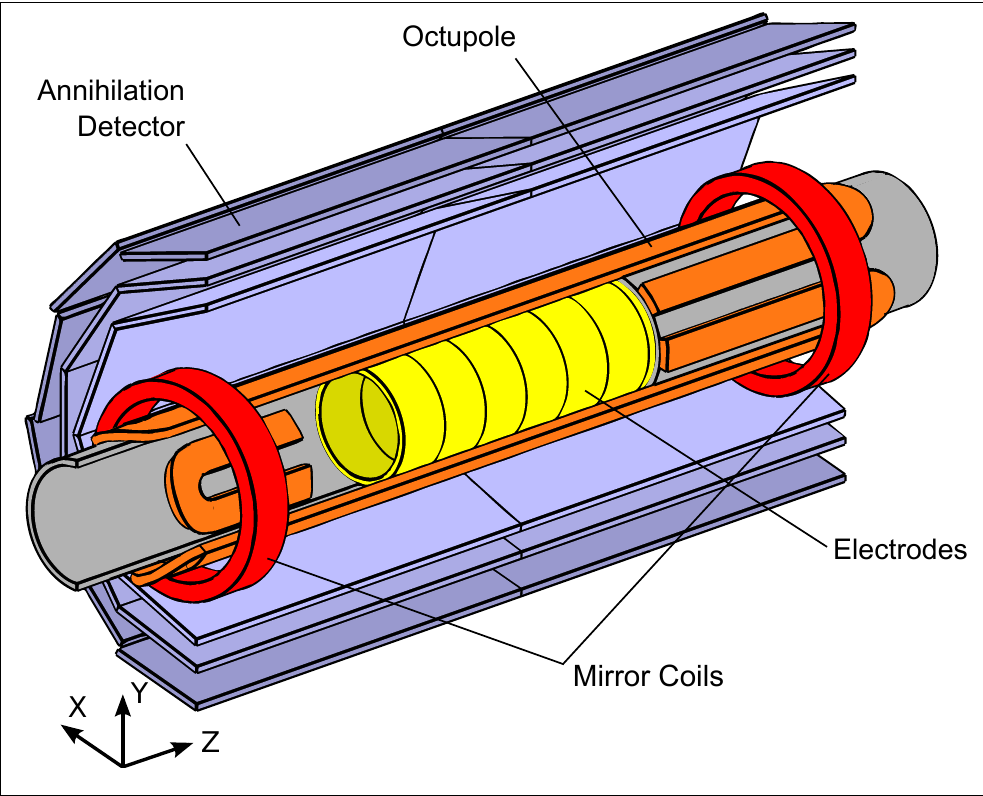}}
\caption{A schematic, cut-away diagram of the antihydrogen production and trapping region of the ALPHA apparatus, showing the relative positions of the cryogenically cooled Penning-Malmberg trap electrodes, the minimum-B trap magnets and the annihilation detector. The trap wall is on the inner radius of the electrodes. Not shown is the solenoid, which makes a uniform field in $\zhat$.  The components are not drawn to scale.}
\label{apparatus}
\end{figure}

Recently, antihydrogen (\Hbar) atoms were trapped in the ALPHA apparatus at CERN \cite{andr:10a,andr:11a}.  The ability to discriminate between trapped antihydrogen and incidentally trapped antiprotons was crucial to proving that antihydrogen was actually trapped \cite{andr:10a,andr:11a,andr:11b}.  The antihydrogen was trapped in a magnetic minimum \cite{prit:83} created by an octupole magnet which produced fields of $1.53\,\mathrm{T}$ at the trap wall at $\Rw=22.28\,\mathrm{mm}$, and two mirror coils which produced fields of $1\,\mathrm{T}$ at their centers at $z=\pm 138\,\mathrm{mm}$.  The relative orientation of these coils and the trap boundaries are shown in Figure~\ref{apparatus}.  These fields were superimposed on a uniform axial field of $1\,\mathrm{T}$ \cite{bert:06,andr:08b}.  The fields thus increased from about $1.06\,\mathrm{T}$ at the trap center ($r=z=0\,\mathrm{mm}$), to $2\,\mathrm{T}$ at the trap axial ends ($r=0\,\mathrm{mm}$, $z=\pm 138\,\mathrm{mm}$), and to $\sqrt{1.06^2+1.53^2}\,\mathrm{T}=1.86\,\mathrm{T}$ on the trap wall at ($r=\Rw$, $z=0\,\mathrm{mm}$). \footnote{Note that $0.06\,\mathrm{T}$ is field from the mirrors at $z=0\,\mathrm{mm}$.}  Antihydrogen was trapped in this minimum because of the interaction of its magnetic moment with the inhomogeneous field. Ground state antihydrogen with a properly aligned spin is a low field seeker; as its motion is slow enough that its spin does not flip, the antihydrogen is pushed back towards the trap center \footnote{Because of the interaction between the mirror and octupole fields, the magnetic field minimum is actually slightly radially displaced from the trap center, not at the trap center itself.} by a force
\begin{equation}
\mathbf{F}=\nabla (\muHbarb\cdot\mathbf{B}),
\label{RestoreForceHbar}
\end{equation}
where $\mathbf{B}$ is the total magnetic field, and $\muHbarb$ is the antihydrogen magnetic moment. Unfortunately, the magnetic moment for ground state antihydrogen is small; the trap depth in the ALPHA apparatus is only $\ETrap=0.54\,\mathrm{K}$, where $\mathrm{K}$ is used as an energy unit.

Trapped antihydrogen was identified by quickly turning off the superconducting octupole and mirror magnetic field coils. Any antihydrogen present in the trap was then released onto the trap walls, where it annihilated.  The temporal and spatial coordinates of such annihilations were recorded by a vertex imaging particle detector \cite{andr:11b,fuji:08,andr:11c}.  The detector is sensitive only to the charged particles produced by antiproton annihilations; it cannot detect the gamma rays from positron annihilations.  Thus, it cannot directly discriminate between antihydrogen and any bare antiprotons that might also be trapped.  We must use additional means to prove that a candidate observation (event) results from an antihydrogen annihilation.

Bare antiprotons can be trapped by the octupole and mirror fields because they may be reflected, or mirrored \cite{chen:84:Mirror}, by the increasing field as they propagate away from  the trap center.  Antiprotons obey the Lorentz force,
\begin{equation}
\mathbf{F}=-q(\mathbf{E}+\mathbf{v}\times\mathbf{B}),
\label{LorentzEquation}
\end{equation}
where $q$ is the unit charge, $\mathbf{v}$ is the antiproton velocity, and $\mathbf{E}$ is the electric field, if any, present in the trap. In our circumstances, the antiprotons generally satisfy the guiding center approximation requirements \cite{litt:81}. Temporarily ignoring $\mathbf{E}$, the force law for the antiprotons reduces to one similar to that for antihydrogen, (\ref{RestoreForceHbar}),
\begin{equation}
\mathbf{F}=\nabla (\mupbarb\cdot\mathbf{B}),
\label{RestoreForcePbar}
\end{equation}
with the antiproton perpendicular magnetic moment $\mupbarb$ replacing $\muHbarb$ in (\ref{RestoreForceHbar}), and with the additional constraint that the antiprotons follow the magnetic field lines, slowly progressing between lines as dictated by the other guiding center drifts.  Here, $\mupbar=|\mupbarb|=\Eperp/B$, $\mupbarb$ is aligned antiparallel to $\mathbf{B}$, and $\Eperp$ is the antiproton kinetic energy perpendicular to $\mathbf{B}$.  Because $\mupbar$ is adiabatically conserved, antiprotons can be trapped if their parallel energy is exhausted as they propagate outwards from the trap center.  The trapping condition comes from the well-known magnetic mirror equation,
\begin{equation}
\Bmax=B_0\left(1+{\frac{\Eparzero}{\Eperpzero}}\right),
\label{MirrorField}
\end{equation}
which defines the largest total magnetic field $\Bmax$ to which an antiproton that starts at the trap center can propagate.  Here, $B_0$ is the total magnetic field magnitude at the trap center, and $\Eperpzero$ and $\Eparzero$ are the antiproton's kinetic energies perpendicular and parallel to the total magnetic field at the trap center.  Using (\ref{MirrorField}), we can readily define the critical antiproton trapping energy ratio
\begin{equation}
\Rpbarc=\frac{B_0}{\Bwall - B_0},
\label{pbarTrapRatio}
\end{equation}
where $\Bwall$ is the smallest total magnetic field magnitude on the region of the trap wall accessible from the trap center.\footnote{More completely, $\Bwall$ is the lesser of the total magnetic field at the trap wall or the total magnetic field on the trap axis
   directly underneath the mirror.  In our case, the former is lower.} An antiproton will be trapped if its ratio of $\Eperpzero/\Eparzero$ exceeds $\Rpbarc$, i.e., if its perpendicular energy is large compared to its parallel energy.

A typical antihydrogen synthesis cycle \cite{andr:10a} starts with 15,000--30,000 antiprotons\footnote{The lower number (15,000) characterizes the number of antiprotons when we employ antiproton evaporative cooling \cite{andr:10}.} trapped in an electrostatic well, and several million positrons trapped in a nearby electrostatic well of opposite curvature. This configuration is called a double well Penning-Malmberg trap; the electrostatic wells provide axial confinement, and the aforementioned axial magnetic field provides radial confinement.  The antiprotons come from CERN's Antiproton Decelerator (AD) \cite{maur:97}, and the positrons from a Surko-style \cite{murp:92} positron accumulator.  The reader is referred elsewhere \cite{andr:10,andr:11b} for the details of the trap operation; we will here discuss only those aspects of the trap operation relevant to discriminating between antiprotons and antihydrogen.

About one third of the antiprotons convert to antihydrogen on mixing with positrons \cite{andr:11a}.  Some of these antihydrogen atoms hit the trap wall and annihilate.  Others are ionized by collisions with the remaining positrons or antiprotons \cite{andr:11b}, or by the strong electric fields present in mixing region \cite{gabr:02,andr:10d}, and turn back into bare antiprotons (and positrons).  Only a very few antiatoms are trapped at the end of the mixing cycle, and confined with these few are approximately 10,000--20,000 bare antiprotons.  If these antiprotons were isotropically distributed in velocity, it is easy to show by integrating over the distribution that the fraction that would be trapped by the octupole and mirror fields alone once the electrostatic fields are turned off is:
\begin{equation}
\frac{1}{\sqrt{\Rpbarc+1}}.
\label{TrappingFraction}
\end{equation}
Since $\Rpbarc=1.35$ for our magnet system, 65\%, of an isotropically distributed population of antiprotons would be trapped.\footnote{This calculation assumes that the antiprotons originate at the magnetic minimum in the trap.  The parameter, $\Rpbarc$, is greater for antiprotons that originate elsewhere, so, for such antiprotons, the trapping fraction would be less.} The actual distribution of the bare antiprotons is unknown and likely not isotropic. Nevertheless, if any fraction of these antiprotons were actually present in the trap when the magnets are shut off, the signal from these antiprotons would overwhelm the signal from any trapped antihydrogen.  Thus, our goal is two-fold: (1) in the experiment, to eliminate the trapped antiprotons if possible, and (2) in the analysis, to be able to discriminate between trapped antihydrogen and any mirror-trapped antiprotons that might have survived the elimination procedures.

In Section~\ref{Simulations} of this paper, we describe the numeric simulations that we used to investigate these issues. In Section~\ref{PbarClearing}, we describe how we apply large electric fields which clear all antiprotons with kinetic energy less than about $50\,\eV$.  In Section~\ref{PbarGeneration} we consider the various mechanisms that could result in mirror-trapped antiprotons with this much energy and conclude that few, if any, antiprotons are trapped.  In Section~\ref{Benchmarking} we describe experiments which benchmark the antiproton simulations, and in Section~\ref{HbarDist} we discuss the postulated antiproton energy distribution.  Finally, in Section~\ref{TrappingExperiments}, we employ simulations to show that if any mirror-trapped antiprotons were to survive the clearing processes, they would annihilate with very different temporal and spatial characteristics than do minimum-B trapped antihydrogen atoms.

\section{Antiproton and antihydrogen simulations}
\label{Simulations}
In this section, we first describe how we calculate the electric and magnetic fields present in the apparatus, including the effects of eddy currents while the magnets are being turned off (shutdown). Then we describe the simulation codes that use these fields to determine the antiproton and antihydrogen trajectories.

\subsection{Fields}
\label{Fields}
Electric fields are generated in the trap by imposing different potentials on the trap electrodes (see Figure~\ref{apparatus}).  In the simulations, these fields are determined by finite difference methods.  Two independent calculations were undertaken.  The first, and the one used in the majority of the simulations, was hand coded and used a slightly simplified model of the electrode mechanical structure; the second was obtained using the COMSOL Multiphysics package \cite{comsol} and an exact model of the electrode mechanical structure.  When the calculations were compared,
the largest differences in the potentials were near the gaps between the electrodes at the trap wall.  These differences reflected the handling of the computational grid near the electrode gaps.  The largest potential energy differences were more than two orders of magnitude smaller than the antiproton energy scale. Away from the electrode gaps, these differences were more than four orders of magnitude smaller. The annihilation location statistics that result from the two finite difference calculations agree within $\sqrt{N}$ fluctuations.

Four magnetic field coils, a solenoid, two mirrors and an octupole, produce the fields modeled in the simulations.  (A fifth coil present in the experiment, a solenoid which boosts the magnetic field during the antiproton catching phase, is not energized during the times studied in the simulations.)  No simple analytic expressions for the field from these coils exist because their windings possess an appreciable cross-sectional area and are of finite length.  Consequently, we use the Biot-Savart numeric integrator found in the TOSCA/OPERA3D field solver package \cite{opera} to generate a three-dimensional magnetic field map \cite{bert:06}.  Granulation issues make the direct use of this map problematic in our particle stepper, so we use the map to find the parameters of an analytic model of the vector magnetic potential, $\mathbf{A}$, from which we then derive the field.  Using this analytic expression for $\mathbf{A}$ is computationally efficient, requires little memory, and eliminates the granulation issues. Over most of the particle-accessible space the fields derived from $\mathbf{A}$ are an excellent match to the numeric fields; the deviation between the numeric and analytic fields is never greater than about 2\%, and is this large only near the axial ends of the octupole where particles rarely reach.  However, while the fields derived from $\mathbf{A}$ satisfy $\nabla\cdot\mathbf{B}=0$ exactly, they do not quite satisfy $\nabla\times\mathbf{B}=0$, and require the existence of unphysical currents, principally near the mirror coils.  These currents are very small; over the majority of the trap, the unphysical current densities are more than four orders of magnitude lower than the typical current densities in the mirror coils. Even near the wall under the mirror coils where the unphysical current densities are largest, they are still more than two orders of magnitude lower than the typical current densities in the mirror coils. To further test the validity of this analytic calculation of $\mathbf{B}$, we studied the distribution of annihilation locations with a computationally slower, but more accurate $\mathbf{B}$ found via the Biot-Savart line integral methodology. Since none of these studies showed statistically significant differences in the antiproton annihilation location distributions, we used the faster analytic calculation of $\mathbf{B}=\nabla\times\mathbf{A}$ throughout this paper. Routines to calculate $\mathbf{A}$ and $\mathbf{B}$ were implemented independently in two different computer languages.  The results of the two implementations were each checked against the numeric field map and against each other.  The details of the calculation of $\mathbf{A}$ are given in \ref{MagneticField}.

An important advantage of the vector magnetic potential formulation is that it makes it trivial to calculate the electric field induced by the decaying magnetic field during the magnet shutdowns.  This electric field, given by $\mathbf{E}=-\partial\mathbf{A}/\partial t$, plays a key role in antiproton dynamics as it is responsible for conserving the third (area) adiabatic invariant \cite{chen:84:Third}.

The steady-state coil currents are measured to 1\% accuracy, and this sets the accuracy to which the fields are known. During the magnet shutdown, the coil currents decay in a near exponential fashion with measured time constants near $9\,\mathrm{ms}$. (In the simulations, we use the measured coil current decays to capture the small deviations from exponential decay.) However, the changing magnetic field induces currents in the trap electrodes which retard the decay of the field.  We have found these decay currents using the COMSOL Multiphysics package \cite{comsol} and a precise model of the electrode mechanical structure.  The eddy currents depend on the resistivity of the 6082 aluminum from which the electrodes are fabricated.  This resistivity is $3.92\times 10^{-8}\,\Omega\mathrm{m}$ at room temperature, and is reduced, at cryogenic temperatures, by the Residual Resistance Ratio (RRR), which we measured to be 3.06.  We find that the eddy currents delay the decay of the magnetic field in a manner well-modeled by passing the coil-created magnetic field through a single-pole low pass filter; the filter time constants are  $1.5\,\mathrm{ms}$ for the mirror coils and  $0.15\,\mathrm{ms}$ for the octupole field. The eddy currents have more influence on the mirror fields than the octupole fields because the breaks between the electrodes do not interrupt the largely azimuthal currents induced by the mirrors, but do interrupt the largely axial currents induced by the octupole.  The simulations employ these filters to model the effects of the eddy currents.

\subsection{Antiproton Simulations}
The antiproton simulations push particles in response to the Lorentz force (\ref{LorentzEquation})
using the fields of Section~\ref{Fields}.  Two codes were developed. The first and primary code propagates the full Lorentz force equations for the position $\mathbf{r}$ and velocity $\mathbf{v}$ using the Boris method \cite{Birdsall1985}
\begin{eqnarray}
\mathbf{r}(t+\frac{\delta t}{2}) &=& \mathbf{r}(t-\frac{\delta t}{2}) + \delta t\mathbf{v}(t)\\
\mathbf{v}(t+\delta t) &=&\mathbf{v}(t) - \frac{ q\delta t}{m}
\{\mathbf{E}[\mathbf{r}(t+\frac{\delta t}{2}), t+\frac{\delta t}{2}]\nonumber\\
&+&\frac{\mathbf{v}(t+\delta t)+\mathbf{v}(t)}{2}
\times\mathbf{B}[\mathbf{r}(t+\frac{\delta t}{2}), t+\frac{\delta t}{2}]\},
\end{eqnarray}
where $\mathbf{r}$ is the antiproton position.  This algorithm is an order $\delta t^3$ method for a single timestep $\delta t$. It conserves the perpendicular energy exactly in a uniform, static field; this is particularly important as the simulations must conserve $\mupbar$ adiabatically.

The second code uses guiding center approximations, including $\mathbf{E}\times\mathbf{B}$, curvature, and grad-B drifts, and propagates particles using an adaptive Runge-Kutta stepper.  The results of the two codes were compared, and no significant differences were observed.  Typical antiproton trajectories are described in \ref{pbarOrbits}.

\subsection{Antihydrogen Simulations}
The antihydrogen simulations pushed particles in response to (\ref{RestoreForceHbar}) in the fields of Section~\ref{Fields}.  Two adaptive Runge-Kutta stepper codes were developed independently and the results compared.  No significant differences were observed.  In addition, the usual convergence tests of the simulation results as a function of the time step were satisfactorily performed.  Similar tests were also performed for the antiproton simulations.  Typical antihydrogen atom trajectories are described in \ref{HbarOrbits}.

\section{Antiproton distribution and clearing}
\label{PbarClearing}
\begin{figure}[tb]
\centerline{\includegraphics{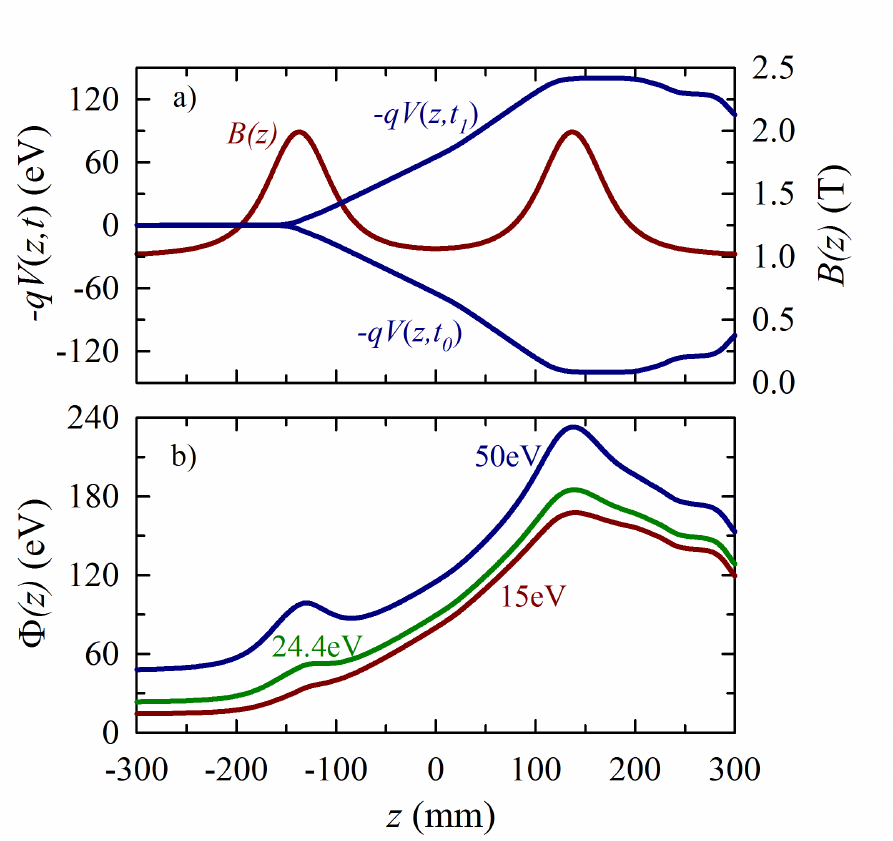}}
\caption{a) The total on-axis magnetic field $B(z)$, and the electrostatic potential energy of an antiproton in the strong clearing fields at times $t_0$ and $t_1$.  b) The pseudopotential (\ref{PseudoPot}) for antiprotons with perpendicular energy $\Eperpzero=15$, $24.4$, and $50\,\mathrm{eV}$.  A well exists in the pseudopotential only for $\Eperpzero>24.4\,\mathrm{eV}$.}
\label{PseudoPotGraph}
\end{figure}

Immediately after a mixing cycle, we axially ``dump'' the antiprotons and positrons onto beamstops where they annihilate.  The dumps use a series of electric field pulses, and are designed to facilitate counting of the charged particles.  They employ relatively weak electric fields. (We switched from an ``original'' dump sequence to an ``improved,'' more efficient, dump sequence midway through the runs reported in this paper.) After the dumps, all the electrodes are grounded; any antiprotons that still remain in the trap must be trapped by the mirror and octupole fields alone. The magnitude of the mirror fields is plotted in Figure~\ref{PseudoPotGraph}a.  Next, we attempt to ``clear'' any such mirror-trapped antiprotons with a series of four clearing cycles.  These clearing cycles use much larger electric fields than the dump pulses; there are two initial ``weak'' clears, and two final ``strong'' clears. The electrostatic potentials used in the strong clears are graphed as $-qV(z,t_0)$ and $-qV(z,t_1)$ in Figure~\ref{PseudoPotGraph}a; the weak clear fields are half as large as the strong clear fields.

A mirror-trapped antiproton can be thought to move in a pseudopotential $\Phi$ which combines the electrostatic potential with an effective potential which derives from the invariance of $\mupbar$,
\begin{equation}
\label{PseudoPot}
\Phi(z,t)= -qV(z,t)+\mupbar B(z).
\end{equation}
For simplicity, we consider $\Phi$ on the $r=0$ axis only.  Figure~\ref{PseudoPotGraph}b plots the pseudopotential for $\mupbar/B_0=\Eperpzero=15\,\mathrm{eV}$, $24.4\,\mathrm{eV}$, and $50\,\mathrm{eV}$.  For an antiproton to be trapped, a well must exist in the pseudopotential.  This condition, which is a function of the perpendicular energy $\Eperpzero$ only, replaces the prior trapping condition, $\Eperpzero/\Eparzero>\Rpbarc$ in the presence of an electric field.  For our parameters, a well only develops for antiprotons with  $\Eperpzero>24.4\,\mathrm{eV}$. Any antiproton with $\Eperpzero<24.4\,\mathrm{eV}$ will necessarily be expelled from the system by the strong clear field even if it has $\Eparzero=0\,\mathrm{eV}$.

\begin{figure}[tb]
\centerline{\includegraphics{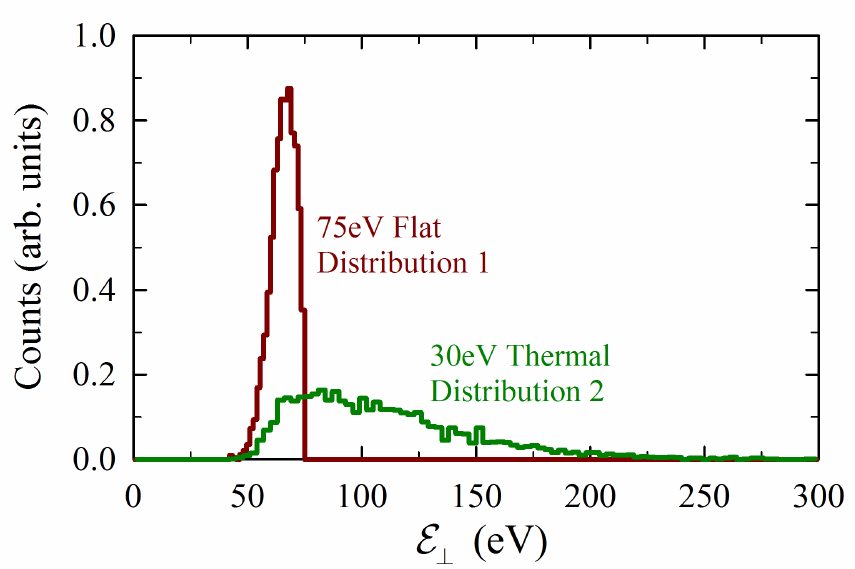}}
\caption{The antiproton distributions that survive the clearing sequence simulations for the initial Distributions 1 and 2 defined in the text.  Other initial distributions (not shown), which, for instance, start all the antiprotons close to the trap axis, yield similar thresholds.}
\label{EperpSurvive}
\end{figure}

It might appear that antiprotons with $\Eperp>24.4\,\mathrm{eV}$ would be trapped. But Figure~\ref{PseudoPotGraph}b shows the static pseudopotential; in the experiment, the clearing field swings from the potential shown in Figure~\ref{PseudoPotGraph}a at time $t_0$ to the potential at time $t_1$ and back eight times (the first four swings, during the weak clears, are at half voltage). Each of these eight stages lasts $12\,\mathrm{ms}$. Extensive computer simulation studies show that these swings expel all antiprotons with $\Eperp<\EperpTrap=50\,\mathrm{eV}$.  Two such studies are shown in Figure~\ref{EperpSurvive}.

The simulations are initiated with a postulated antiproton distribution before the clears.  Unfortunately, we do not know this distribution experimentally (see Section~\ref{PbarGeneration}), so we use two trial distributions that cover the plausible possibilities:  Both distributions assume a spatially uniform antiproton density throughout the trap region but differ in their velocity distribution. Distribution~1 has a velocity distribution that is isotropic and flat up to a total energy of $75\,\mathrm{eV}$, while Distribution~2 has a velocity distribution that is isotropic and thermal with a temperature of $30\,\mathrm{eV}$.  Note that these distributions are intended to reveal the properties of antiprotons that could survive the clears.  They are not intended to be representative of (and, in fact, are thought to be far more extreme than) the actual antiprotons in the trap.

For both distributions, less than 2\% of the antiprotons survive the clearing cycles and remain in the trap, and all that survive have $\Eperp>50\,\mathrm{eV}$.  Further, those with $\Eperp>50\,\mathrm{eV}$ are only trapped if they have very little $\Epar$, as not much parallel energy is needed for them to surmount the relatively shallow pseudopotential wells. For example, in Figure~\ref{PseudoPotGraph}b, the potential well for antiprotons with $\Eperp=50\,\mathrm{eV}$ is only about $12\,\mathrm{eV}$ deep.

The improved efficacy of the time-dependent clearing cycles over the static clearing potential comes from two factors:  1) The repeated voltage swings accelerate the antiprotons, in some cases non-adiabatically.  This often gives them sufficient parallel energy to escape.  2) The potentials depicted in Figure~\ref{PseudoPotGraph}a are generated by voltages impressed on 21 electrodes.  Four central electrodes have a significantly slower temporal response than the outer electrodes;  this creates a momentary well that lifts and eventually dumps antiprotons with increased parallel energy, again raising the likelihood that they escape.

We monitor the antiproton losses in our experiments during the clearing cycles (see table~\ref{ClearSurviveTable}.)  With the original dumps, a substantial number of antiprotons escape in the first clear.  A few antiprotons escape during the second and third clears, but, to the statistical significance of the measurement, none escape in the last clear.  With the improved dumps, far fewer escape in the first clear, a few, perhaps, in the second and third, and none in the last. It is telling that there is no upwards jump in the number that escape between the second and third clears (between the last weak and the first strong clear), as this lack suggests that there is no continuous distribution of antiprotons with a significant population with energies between $\Eperp\approx 25\,\mathrm{eV}$, which are cleared by the weak clears alone, and $\Eperp=50\,\mathrm{eV}$, which are cleared by the strong clears. Thus, in conjunction with the simulations, we conclude that it is not likely that antiprotons with perpendicular energy less than $50\,\mathrm{eV}$ survive the clears and, therefore, none are likely present during the magnet shutdowns.

\begin{table}[tb]
\caption{\label{ClearSurviveTable}The average number of antiproton annihilations detected during the clearing operations. The data includes the false counts from cosmic background, which is separately measured, and given on the last row. The error is the statistical error of the average.  The Trapping rows were measured during normal trapping attempts.  The Benchmarking row was measured while deliberately creating high perpendicular energy antiprotons (see Sec.~\ref{Benchmarking}).  The Full column shows the number of counts observed during the entire $24\,\mathrm{ms}$ time period taken by each clearing cycle.  The Windowed column shows the number of counts between $0.6\,\mathrm{ms}$ and $2\,\mathrm{ms}$ in each cycle.  We know from other data, not shown, that while trapping, almost all the antiprotons escape in this window.  This is expected as it takes $2\,\mathrm{ms}$ for the clearing potentials to reach their peak.  (Employing a $1.4\,\mathrm{ms}$ window increases the signal to noise ratio.) For the Benchmarking trials, antiprotons escape during the entire clearing cycle, and windowing would cut legitimate data. This data was collected by our detector in a non-imaging mode, wherein the detection efficiency is 70\%--95\% assuming most of the antiprotons hit near the trapping region.}
\begin{indented}
\item[]\begin{tabular}{r c c c }
\br
&Full&Windowed&Trials\\
\mr
\makebox[96pt][l]{Trapping \hfill}&&&869\\
\makebox[96pt][l]{~~--Original dumps\hfill}&&&\\
First clear (Weak)   &31.43$\pm$0.21   &31.014$\pm$0.207&\\
Second clear (Weak)  &0.38$\pm$0.02    &0.022$\pm$0.005&\\
Third clear (Strong) &0.37$\pm$0.02    &0.016$\pm$0.004&\\
Fourth clear (Strong)&0.31$\pm$0.02    &0.022$\pm$0.005&\\
\mr
\makebox[96pt][l]{Trapping \hfill}&&&371\\
\makebox[96pt][l]{~~--Improved dumps\hfill}&&&\\
First clear (Weak)   &0.55$\pm$0.04   &0.205$\pm$0.024&\\
Second clear (Weak)  &0.34$\pm$0.03   &0.035$\pm$0.010&\\
Third clear (Strong) &0.33$\pm$0.03   &0.042$\pm$0.009&\\
Fourth clear (Strong)&0.24$\pm$0.03   &0.011$\pm$0.005&\\
\mr
\makebox[96pt][l]{Benchmarking}&&&27\\
First clear (Weak)       &2460$\pm$150& &\\
Second clear (Weak)      &466$\pm$41&   &\\
Third clear (Strong)     &283$\pm$30&   &\\
Fourth clear (Strong)    &45.9$\pm$6.7& &\\
\mr
\makebox[96pt][l]{Background\hfill}&&&\\
&0.32$\pm$0.03       &0.019$\pm$0.002&\\
\br
\end{tabular}
\end{indented}
\end{table}


\section{Mirror-trapped antiproton creation}
\label{PbarGeneration}
In this section we will describe three scenarios that could result in the creation of mirror-trapped antiprotons: creation during the initial capture and cooling of antiprotons from the AD; creation during the mixing of antiprotons into the positrons; and creation by the ionization of antihydrogen. We will show that none of these mechanisms are likely to produce mirror-trapped antiprotons with $\Eperp$ exceeding $50\,\eV$.  However, the calculations are sufficiently uncertain that they cannot guarantee that none are created.  Instead, we rely on two other arguments: (1) As will be discussed in Section~\ref{TrappingExperiments}, the temporal-spatial characteristics of the candidate events are not compatible with mirror-trapped antiprotons.  (2) By heating the positron plasma, we can shut off the production of antihydrogen \cite{amor:02}.  When we do this, we observe essentially no trapped antihydrogen candidates (one candidate in 246 trials, as opposed to 38 candidates in 335 trials in \cite{andr:10a}).  The temperature to which we heat the positrons, approximately $0.1\,\mathrm{eV}$, is negligible compared to the energy scales discussed in this section, and would have no effect on any mirror-trapped antiprotons created.   These experiments are described in \cite{andr:10a} and will not be further discussed here.  Taken together, these arguments allow us to conclude that few, if any, mirror-trapped antiprotons survive to the magnet shutdown stage where they could confound our antihydrogen signal.

\subsection{Creation on capture from the AD}
\label{PbarGeneration_Capture}
The AD \cite{maur:97} delivers a short pulse of $5\,\mathrm{MeV}$ antiprotons to the ALPHA apparatus.  These antiprotons are passed through a thin metal foil degrader, resulting in a broad antiproton energy distribution.  The slowest of these antiprotons are then captured in a $3\,\mathrm{T}$ solenoidal field (eventually reduced to $1\,\mathrm{T}$) by the fast manipulation of the potentials of a $3.4\,\mathrm{kV}$ electrostatic well \cite{gabr:86,andr:08b}.  Once captured, about 50\% of the antiprotons are cooled to several hundred Kelvin by collisions with the electrons in a pure-electron plasma that had been previously loaded into the same well \cite{gabr:89}.  The electrons themselves cool by emitting cyclotron radiation.  The remaining 50\% of the antiprotons do not cool: they are trapped on field lines at radii greater than the outer radius of the electron plasma and, thus, do not suffer collisions with the electrons.  These uncooled antiprotons are removed from the trap by decreasing the trap depth to, ultimately, about $9\,\mathrm{V}$ on the trap axis, corresponding to $30\,\mathrm{V}$ at the trap wall. (The trap depth on the axis is less than at the wall because of the finite length to radius ratio, $20.05\,\mathrm{mm}/22.28\,\mathrm{mm}$, of the trap electrodes.)  As all of these preparatory steps occur before the neutral trapping fields are erected, any antiproton with $\Epar$ exceeding $9~(30)\,\mathrm{eV}$ will escape before the neutral trap fields are erected and, thus, will not be mirror trapped.

In principle, there is a remote possibility that a high perpendicular energy ($\Eperp>50\,\mathrm{eV}$) antiproton might be largely outside the electron plasma, so that it is not strongly cooled, but would have a parallel energy sufficiently low [$<9~(30)\,\mathrm{eV}$] that it could be trapped in the electrostatic well.  Certainly, as shown by SRIM \cite{zieg:85pbar} calculations, a few antiprotons leave the degrader with such skewed energies.  However, the antiprotons must surmount a $\sim 50\,\mathrm{V}$ blocking barrier to enter the well.  An antiproton could be mirror trapped only if (1) it possessed a high initial $\Eperp$ and a high enough $\Epar$ so that it could pass over this barrier, and then (2) underwent one or more collisions that reduced its $\Epar$ to less than $9~(30)\,\mathrm{eV}$ while leaving its $\Eperp$ above $50\,\mathrm{eV}$.  Such an evolution is unlikely to result from collisions with electrons as the antiproton-electron mass ratio requires multiple collisions to effect a significant change to the antiproton energy, and such a collision sequence does not favor a skewed distribution.  Furthermore, we know experimentally that those antiprotons that do not cool quickly essentially never cool; lengthening the cooling time beyond some tens of seconds does not significantly increase the fraction of antiprotons that are cooled.  Thus, it is unlikely that an antiproton would cool just enough to leave it in a state that could be mirror trapped, but not so much that it cools entirely.

Alternatively, the collisions required to leave a mirror-trapped antiproton might be with another antiproton or with a background neutral gas molecule.  The density of these necessarily high radius antiprotons is low, and, if they are to be mirror trapped, their perpendicular and hence total energy is high.  The exact parameters to use in an antiproton-antiproton collision calculation are unknown, but, under any scenario, only a few antiproton-antiproton collisions will take place during the $80\,\mathrm{s}$ cooling time.  For example, for a plausible density of energetic antiprotons of about $10^4\,\mathrm{cm}^{-3}$, the probability that one $500\,\mathrm{eV}$ antiproton would suffer one collision in $80\,\mathrm{s}$ is approximately $10^{-6}$.  Furthermore, only a small fraction of these collisions would leave the antiprotons with the required skewed energy distribution.

The neutral gas density can be estimated from the antiproton annihilation rate, and is on the order of $10^5\,\mathrm{cm}^{-3}$, if, as is likely, the background gas in our cryogenic trap is $\mathrm{H}_\mathrm{2}$.  \footnote{All gases but H, $\mathrm{H}_\mathrm{2}$ and He freeze out; monatomic H is rare, and there is no source of He.}  While this yields an antiproton-neutral collision rate that is higher than that for antiproton-antiproton collisions, the collision rate calculated by extensions of the methods in \cite{cohe:00} and \cite{cohe:02} is on the order of a few tens of microHertz per antiproton, making it unlikely (few $10^{-3}$) that an individual antiproton will suffer a collision that will leave it with the energies required to be trapped.  Individual antiprotons do not suffer multiple collisions with neutrals on the relevant time scale.

\subsection{Creation during mixing}
\label{PbarGeneration_Mixing}
Antihydrogen is generated by mixing antiproton and positron plasmas after the neutral trapping fields are erected. By this point in the experimental cycle, the two species are cold; the antiprotons are at temperatures of less than two hundred Kelvin, and the positrons are at temperatures less than one hundred Kelvin \cite{andr:10a}. The expected number of antiprotons with an energy exceeding $\EperpTrap$ in a thermalized plasma of $N$ particles and temperature $T$ is $N\exp(-\EperpTrap/\kB T)$, where $\kB$ is Boltzmann's constant.  This number is completely negligible for the relevant temperatures.   The antiproton temperature would have to be approximately 200 times greater ( $\sim 3\,\eV$) for there to be an expected value of one or more antiprotons with energy greater than $50\,\eV$ amongst the $\sim$30,000 antiprotons present in one mixing cycle.  Thus, there is no chance that thermalization of the initial antiproton plasma could produce mirror-trapped antiprotons.

\begin{figure}[tb]
\centerline{\includegraphics{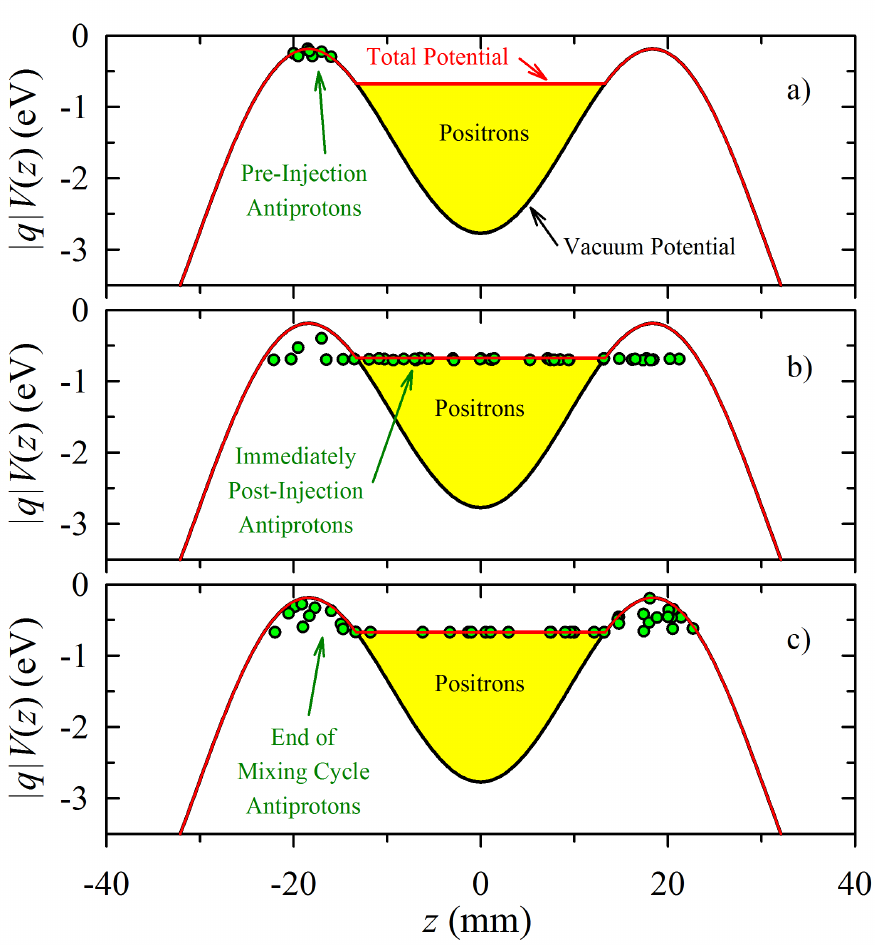}}
\caption{On-axis electrostatic potentials in the mixing region of our apparatus. The green dots are a cartoon depiction of the evolution of the antiprotons. a) Before the autoresonant injection of the antiprotons.  Note how the positron space charge flattens the vacuum potential.  b) Immediately after autoresonant injection of the antiprotons.  c) At the end of the mixing cycle.}
\label{MixingWells}
\end{figure}

During the mixing cycle, the axial motion of the antiprotons is autoresonantly excited \cite{faja:99b,andr:10c} to ease them out of their electrostatic well and into the positron plasma (see Figure~\ref{MixingWells}a--b); the antiprotons phase-lock to a weak, downward-frequency-sweeping oscillating potential applied to a nearby electrode.  The autoresonant drive has a maximum potential drop of $0.05\,\rm{V}$ on the trap axis ($0.1\,\rm{V}$ at the wall), and there are approximately $\Nauto=300$ drive cycles.  Naively, one might think  that there are enough cycles that the drive could excite antiprotons up to the maximum confining potential of $21\,\mathrm{V}$ on the trap wall.  In reality, the antiprotons phase-lock at near $90^{\circ}$ such that the impulse conveyed to the antiprotons on each cycle is small \cite{faja:01a}. The typical antiproton gains just enough energy to enter the positrons: about $0.5\,\mathrm{V}$ on the trap axis when plasma self-fields are included.  If, as occasionally happens, an antiproton loses phase-lock, it will gain a limited amount of additional energy stochastically in rough proportion to $\sqrt{\Nauto}$. Further, this is axial energy; if the antiproton were to somehow gain more than $21\,\mathrm{eV}$ it would be lost immediately unless it had also experienced a sufficient number of collisions to posses substantial perpendicular energy. Under no scenario can the antiproton gain energy close to $50\,\eV$ of perpendicular energy directly from the autoresonant drive.

The autoresonant process injects most of the antiprotons into the positrons, but some are left in the original side well with axial energies up to the electrostatic well depth of about $0.5\,\mathrm{V}$ near the trap axis.  As mixing progresses, antiproton-antiproton collisions cause additional antiprotons to fall into this side well and into the electrostatic well on the other side of the positron plasma (see Figure~\ref{MixingWells}c).  As there is no direct mechanism to transport these antiprotons radially outward \cite{andr:09aa}, most will remain at or near their original radius (between $0.4$ and $0.8\,\mathrm{mm}$ depending on the details of the procedures in use at the time.)   Approximately 50\% of the particles eventually fall into the two side wells, so the number of antiprotons in the side wells eventually approaches the un-mixed antiproton number.  Measurements on similar plasmas show that they thermalize in times on the order of the one second that the mixing continues \cite{andr:10c}. (Unlike in Section~\ref{PbarGeneration_Capture}, the density of these near-axis antiprotons is relatively high.) Measurements also show that evaporative cooling will set their temperature to several times less than the well depth \cite{andr:10}.  Thus, the near-axis antiproton temperature in the side wells is considerably less than $0.5\,\eV$.  The expected number of antiprotons in such a plasma having a perpendicular energy greater than $50\,\eV$ is negligible.

\subsection{Creation by ionization of antihydrogen}
Antihydrogen in the ALPHA experiment is believed formed largely by three body recombination.  This process creates the atoms in highly excited states that can be ionized by sufficiently strong electric fields \cite{gabr:02,andr:10d}.  The strongest electric fields in our trap are found close to the trap wall at the electrode boundaries, and can be as large as $\Emax=42\,\mathrm{V}/\mathrm{mm}$.\footnote{Very close to the electrode gaps, the electrode corners will increase the field beyond $\Emax=42\,\mathrm{V}/\mathrm{mm}$.  However, any antiproton born close enough to corners to feel this enhancement will almost surely hit the wall immediately.} A newly ionized antiproton will be accelerated by these fields, and can pick up perpendicular energy.  However, a careful map of the electric and magnetic fields over the entire trap shows that the perpendicular energy gain cannot exceed more than $3\,\eV$ before the antiproton settles into its $\mathbf{E}\times\mathbf{B}$ motion, so this process cannot lead to mirror-trapped antiprotons.

The arguments in the two previous subsections strongly suggest that antihydrogen cannot be born with substantial center-of-mass kinetic energy under our experimental conditions.  If an antihydrogen atom were, nonetheless, somehow born with high kinetic energy, this energy would be conveyed to the antiproton upon ionization.  Naively, this could lead to a mirror-trapped antiproton.  However, there is an upper limit to the amount of energy an antiproton could possess after ionization.  The limit comes from the Lorentz force equation (\ref{LorentzEquation}). A particle moving at velocity $v$ perpendicular to a magnetic field $\mathbf{B}$ feels a force that is equivalent to that from an electric field of magnitude $vB$.  This magnetic force $qvB$ can ionize an antiproton just as can an electric force $qE$. Thus, if an antihydrogen atom is sufficiently excited that it can be ionized by the large electric field of strength $\Emax$ or less near the trap wall, it will always be ionized by passage through the magnetic field at the center of the trap where it is created if it is moving faster than approximately $\Emax/B_0$.  This sets a rough upper limit on the maximum kinetic energy that a high radius, newly ionized antiproton can have of less than $10\,\eV$.  If an antihydrogen atom has more kinetic energy, it will either: 1) be in a relatively low excited state such that it will not be ionized at all, and will hit the trap wall and annihilate promptly. Or (2) be in an ionizable state and be ionized close to the trap axis by the magnetic force, where it will be thermalized and cooled by the abundant population of antiprotons and positrons found there.  A more exact calculation, given in \ref{Bstrip}, lowers this bound substantially for most antiatoms.

The side wells near the trap wall are as deep as $21\,\mathrm{V}$.  An antiproton that fell into one of these side wells, either indirectly by ionization or directly by some unknown process during mixing, could pick up substantial parallel energy.  However, the density of antiprotons is very low at large radii, and antiproton-antiproton collisions proportionally infrequent.  Multiple collisions would be required to transform the maximum parallel energy of $21\,\eV$ into perpendicular energy of more than $50\,\eV$. Collisions with neutrals, of course, can only lower the antiproton energy. Thus, we can conclude the parallel energy possessed by an antiproton cannot be converted into sufficient perpendicular energy to lead to mirror trapping.

\section{Antiproton simulation benchmarking during magnet shutdowns}
\label{Benchmarking}
The arguments in the previous sections suggest that there are few, if any, mirror-trapped antiprotons. This tentative conclusion relies on information gleaned from the simulations of the efficacy of the clearing cycles. Ultimately, though, we rule out the existence of mirror-trapped antiprotons by comparing their simulated post magnet shutdown dynamics to our experimental observations.  A direct, independent test of the simulations powerfully buttresses our conclusions.

\begin{figure}[tb]
\centerline{\includegraphics{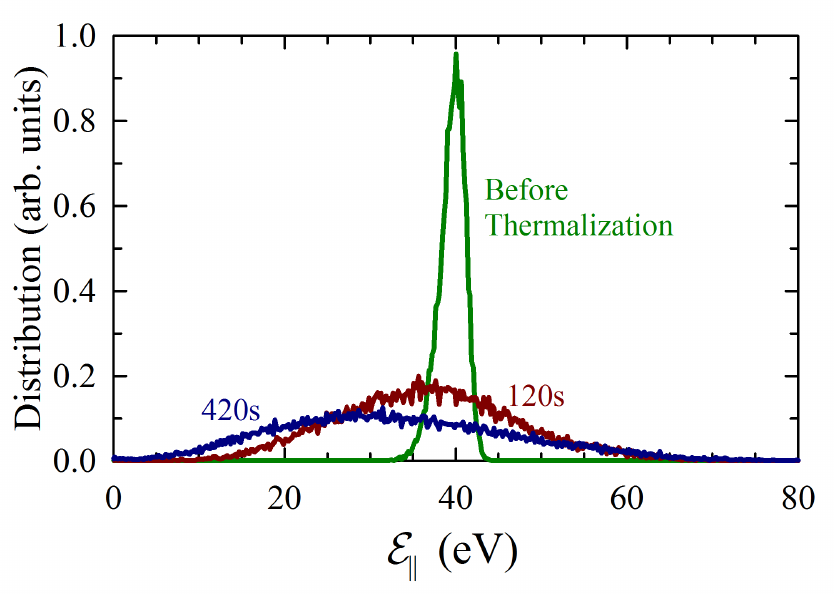}}
\caption{The $\Epar$ distribution of the deliberately-created mirror-trapped antiprotons, shown just after the antiprotons were dropped over the $40\,\mathrm{V}$ potential barrier, at $120\,\mathrm{s}$ post drop, and when thermalization was ended at $420\,\mathrm{s}$ post drop. These distributions were obtained by slowly lowering one of the confining electrostatic barriers, and measuring the number of escaping antiprotons as a function of the barrier height. }
\label{MirTrapDist}
\end{figure}

We performed such a test by deliberately creating a population of mirror-trapped antiprotons.  We began by capturing approximately 70,000 antiprotons from the AD.  These antiprotons were injected over a potential barrier into a deep well, giving them a parallel energy $\Epar$ of approximately $40\,\mathrm{eV}$.  Then the antiprotons were held for $420\,\mathrm{s}$, during which time collisions partially thermalized the populations, transferring parallel energy into perpendicular energy $\Eperp$.  The mean antiproton orbit radius also expanded during this time to approximately $1.5\,\mathrm{mm}$.  Initial, intermediate, and final $\Epar$ distributions are shown in Figure~\ref{MirTrapDist}.  We have no independent measure of the $\Eperp$ distribution.

After the thermalization period, the octupole and mirror coils were energized, followed by the removal of the electrostatic well that had been confining the antiprotons.  Once this well was removed, the antiprotons remaining in the system must have been mirror trapped.  However, many of these antiprotons were not deeply mirror trapped ($\Eperp<\EperpTrap=50\,\mathrm{eV}$), and, as can be seen in ``Benchmarking'' grouping in table~\ref{ClearSurviveTable}, many were expelled during the clears.

\begin{figure}[tb]
\centerline{\includegraphics{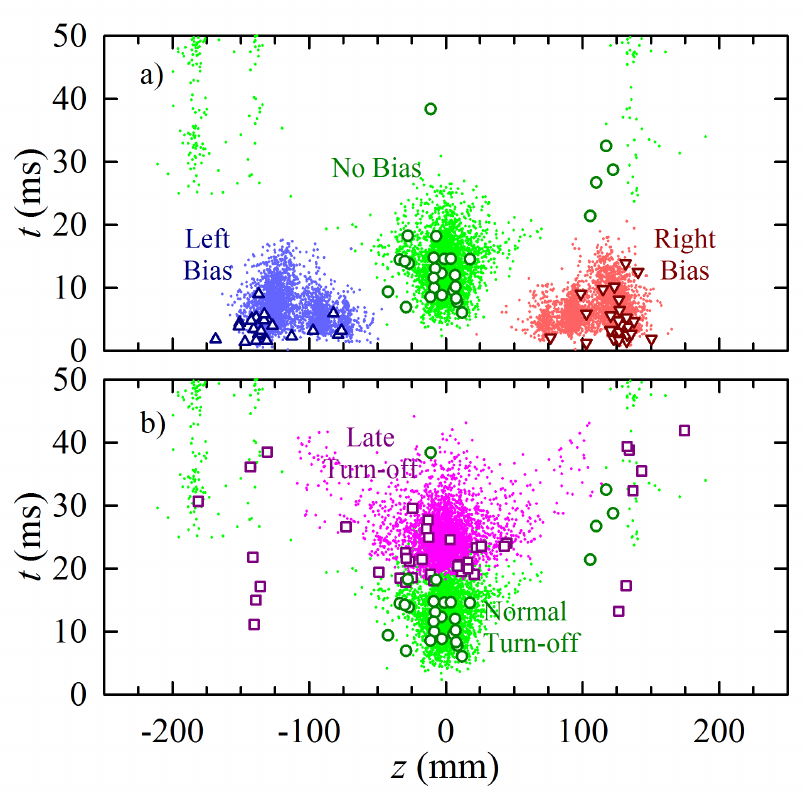}}
\caption{Comparison of the $z$--$t$ annihilation locations of mirror-trapped antiprotons (symbols) and the antiproton simulations (dots).  a) Compares ``Left Bias'' (blue upward pointing triangles and blue dots), ``No Bias'' (green circles and green dots), and ``Right Bias'' (red downward pointing triangles and red dots) for normal current decay times.  b) Compares the ``No Bias'' dataset in a) with normal shutdown timing, to a No Bias dataset in which the octupole decay onset was slowed by about $7\,\mathrm{ms}$ (purple squares and purple dots) relative to the mirror decay onset.   The annihilations near $z=-183\,\mathrm{mm}$, and $\pm 137\,\mathrm{mm}$ are at radial steps in the trap wall.  The detector resolution was approximately $5\,\mathrm{mm}$ in $z$, and $100\,\mu\mathrm{s}$ in $t$; the simulation points were randomly smeared by these resolutions.}
\label{BenchmarkPlot}
\end{figure}

\begin{figure}[tb]
\centerline{\includegraphics{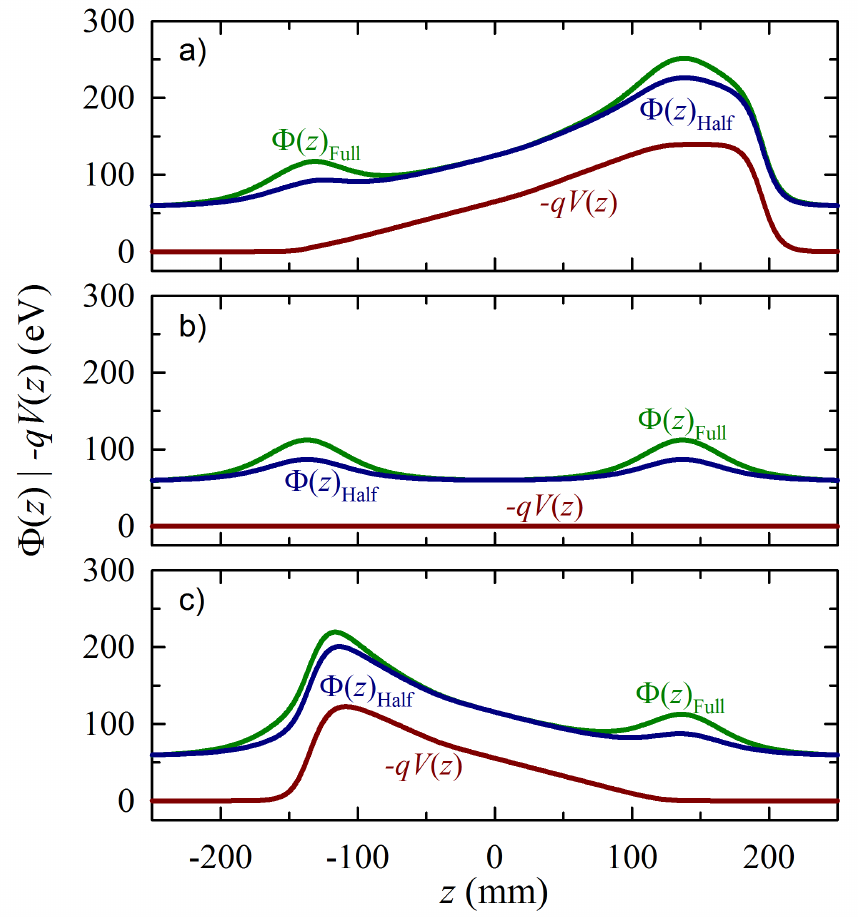}}
\caption{The electrostatic Bias potentials $-qV(z)$, and the on-axis pseudopotentials $\Phi(z)$ for $\Eperp=60\,\mathrm{eV}$ for full and half strength mirror fields and for the a) Left Bias, b) No Bias, c) Right Bias cases.  When the Bias fields are applied, the antiprotons are localized at the ends of the trap.  The localization is preserved as the magnets lose strength during the magnet shutdown.}
\label{QWP_Potentials}
\end{figure}

After the clears, the magnets were turned off, and the annihilation times $t$ and positions $z$ of the remaining antiprotons were recorded.  The results of 27 of these cycles are shown in Figure~\ref{BenchmarkPlot}a.  During most of these cycles, a Bias electric field (see Figure~\ref{QWP_Potentials}) was applied during the magnet shutdown whose intent was to aid the discrimination between bare antiprotons and antihydrogen; the charged antiprotons should be pushed by the Bias field so that they preferentially annihilate on the right side (Right Bias) or on the left side (Left Bias) of the trap, while the uncharged antihydrogen atoms should be unaffected by the Bias field.  In addition, the Bias fields make the pseudopotential wells shallower, so the antiprotons escape and annihilate sooner than when no bias is applied. Figure~\ref{BenchmarkPlot}b shows the effect of delaying the octupole shutdown onset by about $7\,\mathrm{ms}$ relative to the mirror shutdown onset.

Also plotted in Figure~\ref{BenchmarkPlot} are the results of simulating 3364 post-clear survivors.   Since we can only characterize the pre-clear and magnet shutdown antiprotons imperfectly (see Figure~\ref{MirTrapDist}), we must make an estimate for the distribution to use in the simulation.  We believe Distribution~2, defined in Section~\ref{PbarClearing}, is most appropriate as it has a plausible temperature and no strict upper bound on $\Eperp$.   Figure~\ref{BenchmarkPlot} shows that simulations match the experimental data well. Thus, we can confidently use the antiproton simulations as a tool to aid in the discrimination between mirror-trapped antiprotons and antihydrogen. These tests also confirm that the bias fields work as expected. The antihydrogen simulation uses the same magnetic field model as the antiproton simulation, so we have benchmarked the field component of the antihydrogen simulation as well.

\section{Postulated Antihydrogen Energy Distribution}
\label{HbarDist}
As described in Section~\ref{PbarGeneration_Mixing}, antihydrogen atoms are created by mixing antiprotons with positrons.  Initially, the antiprotons have more kinetic energy than the positrons, but we estimate that the antiprotons come into thermal equilibrium with the positrons before the recombination occurs. The positron density is $5\times 10^7\,\mathrm{cm}^{-3}$ and the positron temperature is $40\,\mathrm{K}$ \cite{andr:11a}. We use \cite{hurt:08} to compute a slowing rate of $\sim 200\,\mathrm{s}^{-1}$. From \cite{robi:04a}, the three body recombination rate is approximately $0.1\,\mathrm{s}^{-1}$, but this is the steady state rate to reach a binding of $8\kB T$. Because antihydrogen atoms that have a binding energy of $1 \kB T$ will mostly survive the fields of our trap, the recombination rate will be approximately ten times higher. This is in approximate agreement with our measurements.  Consequently, we expect that the antiprotons cool to the positron temperature before forming antihydrogen.

Because the positron mass is negligible compared to the antiproton mass, a newly formed antihydrogen atom inherits its center-of-mass kinetic energy from the antiproton from which it is formed. Thus, we expect that the antihydrogen itself is in thermal equilibrium with the positrons, and possesses the same distribution function---except that the trapped antihydrogen distribution function is truncated at the energy of the neutral trap depth, $\ETrap=0.54\,\mathrm{K}$.  The positron temperature is much greater than this energy.  Consequently, we expect that the velocity space distribution function $f(v)$ is essentially flat over the relevant energy range for the trapped antihydrogen atoms, and the number of atoms in some velocity range $dv$ is $f(v)v^2\,\mathrm{d}v\propto v^2\,\mathrm{d}v\propto \sqrt{\Energy}\,\mathrm{d}\Energy$.  The number of atoms trapped should be proportional to $\ETrap^{3/2}$.  Note that because $f(v)$ is essentially flat in the relevant region, the antihydrogen distribution $v^2\mathrm{d}v$, once normalized, does not depend in any significant way on the temperature of the positrons.   However, for concreteness, we did our principal antihydrogen simulations with a temperature of $54\,\mathrm{K}$.

\begin{figure}[tbh!]
\centerline{\includegraphics{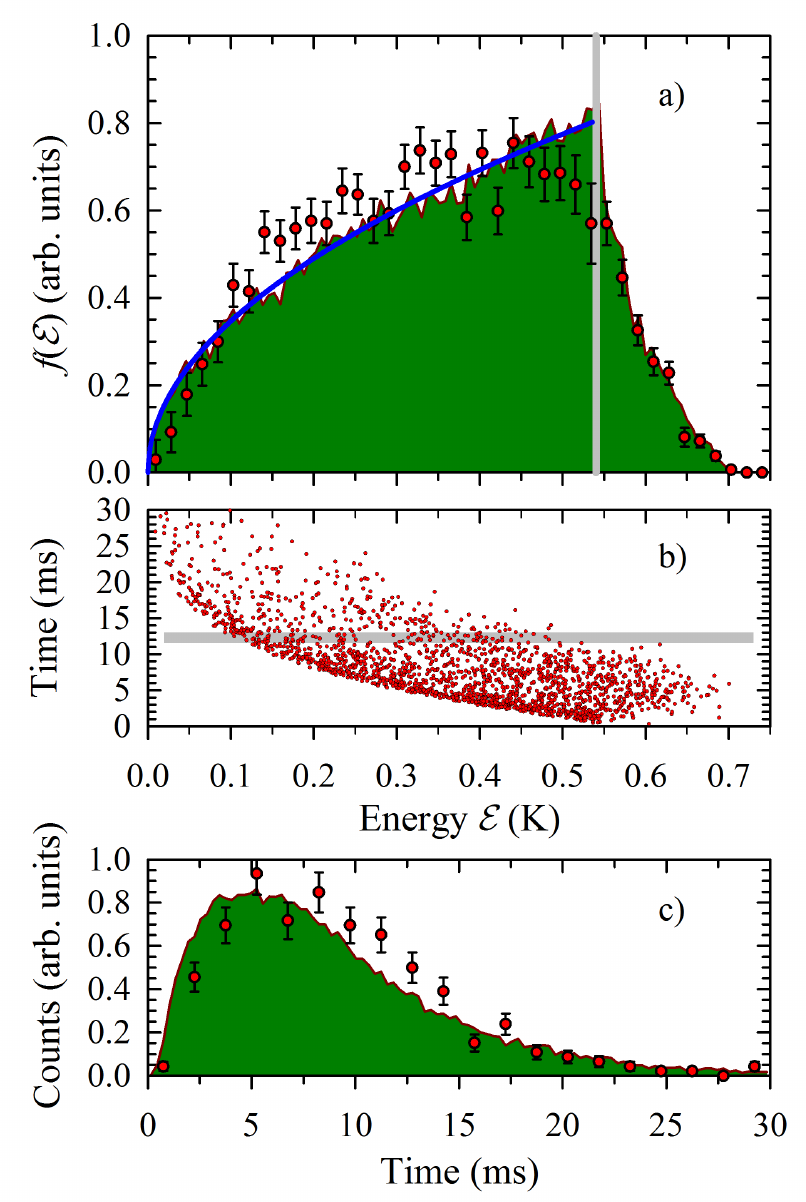}}
\caption{a) The antihydrogen energy distribution $f(\Energy)$.  The solid green area is a histogram of the energy of the trapped antihydrogen atoms as predicted by the simulation from a starting population of atoms at $54\,\mathrm{K}$.  The blue line plots the expected $\Energy^{1/2}$ dependence up to an energy of $0.54\,\mathrm{K}$.   This line ends at the vertical gray line, past which point all the antihydrogen atoms are quasi-trapped. The red points plot the energy distribution function reconstructed from the observed data.  The reconstruction process is discussed in \ref{EnergyReconstruction}. The error bars come from Monte Carlo simulations of the reconstruction process and represent only the statistical errors. b) The time of annihilation after the magnet shutdown as a function of the initial energy for simulated antihydrogen atoms. (For clarity, only a representative 2000 point sample of the 35000 simulated antiatoms is plotted.) The function of the gray band is described in \ref{EnergyReconstruction}. c) Histograms of the number of annihilations as a function of time after the magnet shutdown, as observed in the experiment (red points) and in the simulation (solid green area.)  The error bars on the experimental points come from counting statistics.    }
\label{EnergyDist}
\end{figure}

The simulations reveal that the energy distribution is not strictly truncated at the trapping depth (see Figure~\ref{EnergyDist}a) \cite{andr:11a}.  There exist ``quasi-trapped'' stable trajectories with energies up to about $0.65\,\mathrm{K}$; similar trajectories exist in neutron traps \cite{coak:05}. Quasi-trapped trajectories exist because the antiatom motion is three dimensional.  Rarely is all of the antiatom's motion parallel to the gradient of $|\mathbf{B}|$ at the orbit reflection points at high $|\mathbf{B}|$.  Any motion perpendicular to $\nabla|\mathbf{B}|$, and the kinetic energy associated with this perpendicular motion, is not available to help penetrate through the reflection point.   Hence, the antiatom may be confined even if its energy exceeds the maximum trapping depth. Being only quasi-trapped, these antiatoms are more susceptible to  perturbations than antiatoms trapped below the trapping depth.  We do not know if the quasi-trapped trajectories are long-term stable.

The positron plasma is Maxwellian in the frame that rotates with the positron plasma. This rotation modifies the lab frame distribution. If the positron density were to be very high, the rotation would impart significant additional kinetic energy to the antiprotons, and, hence to the resulting antiatoms. This would result in fewer antiatoms being caught in the trap.  For our densities and fields, however, this effect is small. The reduction in the number of antiatoms that can be trapped from this effect is less than $5\%$.

\section{Trapping Experiments}
\label{TrappingExperiments}
During the 2010 experimental campaign, we observed 309 annihilation events compatible with trapped antihydrogen.  These events were observed under a number of different conditions, including runs with Left Bias, No Bias, and Right Bias, and with the antihydrogen held for times ranging from $172\,\mathrm{ms}$ to $2000\,\mathrm{s}$.  The conditions under which the observed events were obtained are listed in table~\ref{TrapEvents}.

\begin{table}[tb!]
\caption{\label{TrapEvents} Observed trapping events during the 2010 experimental campaign. ``Hold Time'' denotes the time interval between when most of the antiprotons were dumped from the trap, thereby ceasing antihydrogen synthesis, and when the trap magnets are turned off; i.e.\ the approximate minimum time that the antihydrogen was trapped.  As the trapping rate improved continuously during the experimental campaign in 2010, and long Hold trials were all clustered near the end of the campaign, no conclusions about the lifetime of antihydrogen in our trap can be reached from from the ratio of observed trapping events to the number of trials \cite{andr:11a}.}
\begin{indented}
\item[]\begin{tabular}{r|r r r|r r}
\br
Hold&Left Bias&No Bias&Right Bias&Total&Trials\\
Time (s)&&&&&\\
\mr
0.2     &73   &41  &13  &127   &613 \\
0.4     &129  &    &17  &146   &264 \\
10.4     &6    &    &    &6     &6 \\
50.4     &4    &    &    &4     &13\\
180.4    &10   &    &4   &14    &32 \\
600.4    &     &    &4   &4     &38 \\
1000.4   &5    &    &2   &7     &16 \\
2000.4   &     &    &1   &1     &3 \\
3600.4   &     &    &    &      &1 \\
\mr
Total     &227  &41  &41  &309   &  \\
Trials    &577  &227 &182  &      &985  \\
\br
\end{tabular}
\end{indented}
\end{table}

\begin{figure*}[tb!]
\centerline{\includegraphics{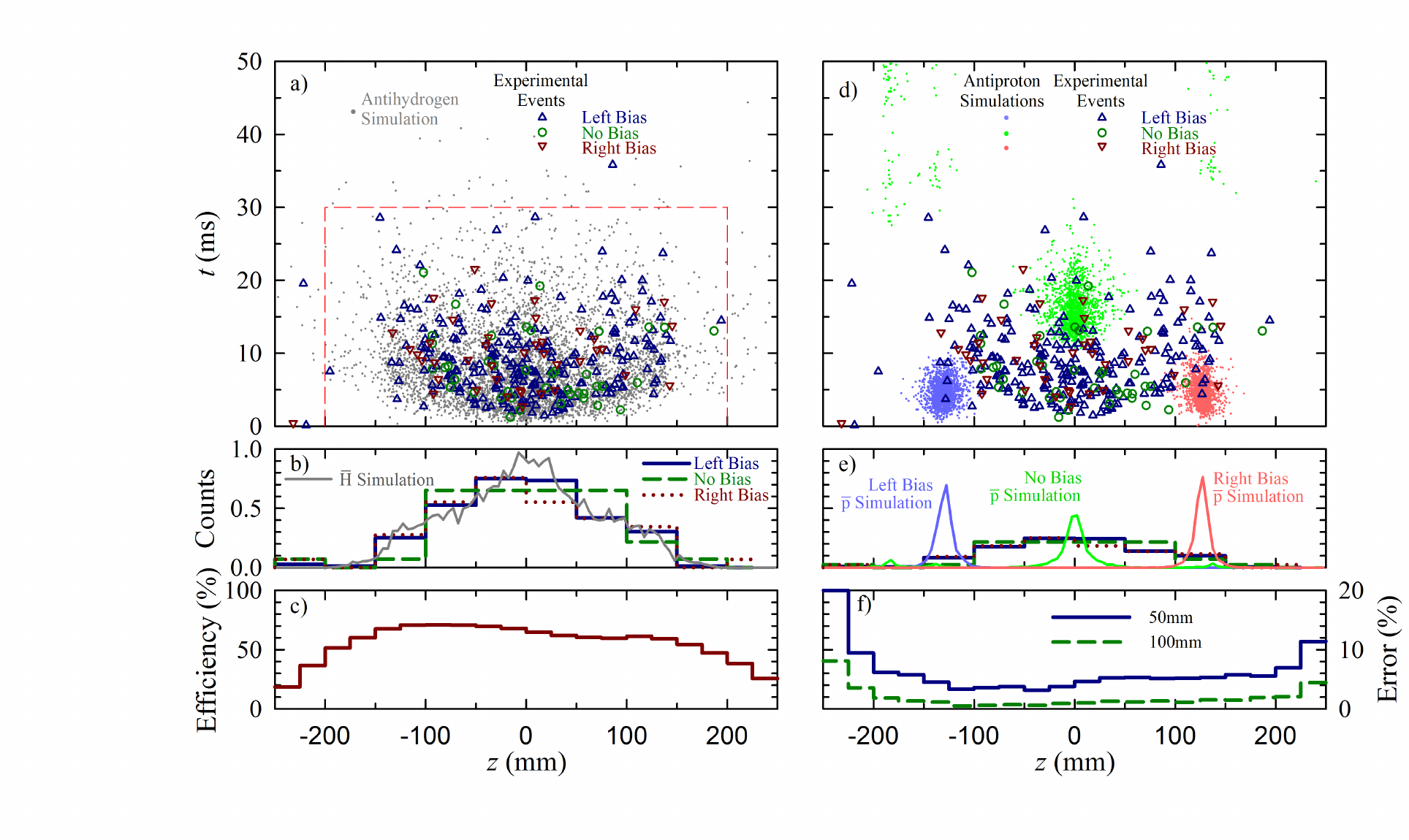}}
\caption{a) Spatial and temporal ($z$--$t$) locations of the annihilations during the trapping events, and the annihilation coordinates predicted by the antihydrogen simulations (small gray dots).  Table~\ref{TrapEvents} details the trapping conditions.  b) $z$ histograms of the annihilation locations. The observed locations agree well with the predictions of the antihydrogen simulation, and are independent of the Bias conditions. c) Detector efficiency as a function of $z$, as calculated by GEANT~3 \cite{brun:87}. d) Similar to a), but with the annihilations predicted by the antiproton simulations for Left Bias conditions (left clump of purple dots), No Bias conditions (central clump of green dots) and Right Bias conditions (right clump of red dots).  e) Similar to b), but with histograms from the antiproton simulations. The counts in the simulation histograms are divided by a factor of five so that the observed event histogram is also visible.) f) Percentage of the reconstructions that are more than $50\,\mathrm{mm}$ and $100\,\mathrm{mm}$ from their true position, as calculated by GEANT~3.}
\label{TrappingAnnihilations}
\end{figure*}

Figure~\ref{TrappingAnnihilations} plots the spatial and temporal ($z$--$t$) locations of the observed annihilations after the octupole and mirror fields were turned off.  The figure compares the observed annihilation locations to the locations predicted by the antihydrogen and antiproton simulations.  The initial distributions in these simulations were the flat antihydrogen distribution discussed in Section~\ref{HbarDist}, and the antiproton Distribution~1 defined in Section~\ref{PbarClearing}.  We chose Distribution~1 here rather than Distribution~2, because we wanted to maximize the number of antiprotons just above the mirror-trapping barrier $\EperpTrap$; for the benchmarking test in Section~\ref{Benchmarking}, we chose Distribution~2 because we had independent evidence (Figure~\ref{MirTrapDist}) of the existence of antiprotons well above $\EperpTrap$.  However, as is evident from comparing Figure~\ref{BenchmarkPlot}a and Figure~\ref{TrappingAnnihilations}d, the differences between the annihilation locations for these two distributions are minor; principally, some of the higher energy antiprotons in Distribution~2 annihilate closer to the center of the trap than the antiprotons in Distribution~1.

In general, the agreement between the observed events and the antihydrogen simulations is excellent; in contrast, the vast majority of observed events are incompatible with the antiproton simulations. As expected, the locations of the observed events are independent of the bias electric fields, as they are in the antihydrogen simulations.  The simulations show, however, that the annihilation locations of postulated antiprotons are strongly dependent on the bias fields. Other simulations, not shown here, show that these conclusions remain true in the face of antiproton energies up to several $\mathrm{keV}$ and gross magnetic field errors.

In ~\cite{andr:10a, andr:11a,andr:11b} we limited our analysis to annihilations which occur within $30\,\mathrm{ms}$ of the beginning of the magnet shutdown.  This criterion was based on the observation that, in the simulations, 99\% of the antihydrogen atoms annihilated by $30\,\mathrm{ms}$.  Here, and in \cite{andr:11a}, we impose an additional requirement that $|z|<200\,\mathrm{mm}$; beyond this region, the efficiency of the detector falls and the accuracy of the detector reconstructions becomes suspect (see Figure~\ref{TrappingAnnihilations}c,f).  In the first $50\,\mathrm{ms}$ after the shutdown, we observed four events which fail these cuts, and, thus, do not appear to be antihydrogen atoms.  These events also appear to be incompatible with the antiproton simulations.  While we have no definitive explanation of these events, there are several possibilities: 1) Even if all 309 events were genuinely due to antihydrogen, we would expect 1\% of the events to be improperly excluded because of the $t$ criterion; the total number of events thereby improperly excluded would be expected to be three.  2) As discussed in ~\cite{andr:10a, andr:11b}, cosmic rays are miscategorized as antiproton annihilations at a rate of approximately $47\,\mathrm{mHz}$.  The events discussed here were observed in approximately $985\times 50\,\mathrm{ms}\approx 50\,\mathrm{s}$, so we would expect to observe approximately 2 such miscategorized cosmic rays, some of which could occur outside the cut boundary.  3) The basic $z$ resolution of our detector is approximately $5\,\mathrm{mm}$, but there is a low probability long tail of badly resolved annihilations (Figure~\ref{TrappingAnnihilations}f).  Some of these observed events may be outside the $|z|<200\,\mathrm{mm}$ window because they were poorly resolved.  4) The trap electrodes have offsets of up to about $50\,\mathrm{mV}$ due to the non-ideal behaviour of the electrode amplifiers.  This creates shallow wells, which might store antiprotons outside of the region in which the clearing fields are applied, and which might cause antiprotons to be released at odd times and positions.

As remarked above, annihilations typically occur within $30\,\mathrm{ms}$ of the magnet shutdown.  The time history of these annihilations contains information about the energy distribution of the antihydrogen atoms \cite{andr:11a}.  Figure~\ref{EnergyDist}b plots the annihilation time as a function of energy as found in the simulations.  As expected, the higher energy antiatoms, which are freed at higher values of the diminishing trap depth, annihilate sooner than low energy antiatoms.  Figure~\ref{EnergyDist}c shows a histogram of the expected and observed annihilation times.  The observed points are well predicted by the simulations.  From the data in Figure~\ref{EnergyDist}c, the original energy distribution of the antiatoms can be coarsely reconstructed, as shown in Figure~\ref{EnergyDist}a.  To within the predictive power of the reconstruction, the energy distribution follows the expected $\Energy^{1/2}$ plus quasi-trapped distribution.   The reconstruction algorithm, and its very significant limits, are described in \ref{EnergyReconstruction}. The influence of the energy distribution on the $z$ distribution is described in \cite{andr:11a}.

\section{Conclusions}
We have presented a detailed study of the behaviour of antihydrogen atoms and antiprotons confined in a magnetic minimum trap. This study was used to guide experiments which eliminate antiprotons as a possible background in recent antihydrogen trapping experiments.  We have demonstrated how the very different behaviour of the neutral and charged particles lead to very different loss patterns in time and space when the magnetic minimum trap is rapidly de-energized. These different loss patterns have been a crucial factor in the identification of trapped antihydrogen. Finally, we have shown how we can use the simulations to reconstruct the energy distribution of the trapped antihydrogen from the time history of the loss after de-energization. These studies and tools have provided important insights into the nature of antihydrogen trapping dynamics.

In the future, it may be possible to discriminate between antihydrogen and antiprotons via a resonant interaction with the atomic structure of the antiatoms.  Such resonant interactions could photoionize the antiatoms or flip their spins such that the antiatoms become high field seekers.  The technique of field-ionization, which has been successfully used with excited antiatoms \cite{gabr:02},  does not work with ground state antiatoms because the fields required to strip a ground state antihydrogen atom are too large, and, thus, would not detect the long-trapped atoms discussed here \cite{andr:11a}. Until efficient resonant interactions  with the antiatoms can be obtained, the techniques demonstrated in this paper will remain a crucial tool in the endeavour to increase the trapping rates and pursue the path towards detailed spectroscopy of antihydrogen.

This work was supported by CNPq, FINEP/RENAFAE (Brazil), ISF (Israel), MEXT (Japan), FNU (Denmark), VR (Sweden), NSERC, NRC/TRIUMF, AITF, FQRNT (Canada), DOE, NSF (USA), and EPSRC, the Royal Society and the Leverhulme Trust (UK).

\appendix

\section{Magnetic Field Formulas}
\label{MagneticField}
In this Appendix, we develop the analytic model of the magnetic field referred to in Section~\ref{Fields} and required for use with the simulations.
\subsection{Mirror coils}
 By comparison with the precise Biot-Savart fields, we found that the magnetic field from each individual mirror coil could be accurately approximated using a pair of circular loops. From Jackson \cite{jack:99a1}, the vector
potential from a single loop is
\begin{equation}
\label{ExactLoop}
A_\phi = C\frac{s}{(r^2+a^2)^{3/2}}(1 +\frac{15}{8}\frac{a^2s^2}{(r^2+a^2)^2}+...),
\end{equation}
where $\phi = \arctan (y/x)$,
$C=I\mu_0a^2/4$ is a constant, $a$ is the radius of the loop,
$s^2=x^2+y^2$, and $r^2=s^2+z^2$ with $x,y,z$ measured from the
center of the circle defined by the loop. Unfortunately, the series
converges very slowly near the mirror and this formula, although accurate,
was abandoned.  Instead, we used a method based on guessing a form for $\mathbf{A}$. The guess
is inspired by the form of the exact $\mathbf{A}$ from a single loop:
\begin{equation}
A_\phi = C \frac{1}{2a\lambda}[(a^2+r^2-2a\lambda s)^{-1/2}-
(a^2+r^2+2a\lambda s)^{-1/2}],
\end{equation}
where all of the parameters are as before and $\lambda$ is a dimensionless fit parameter. Note, the choice $\lambda = \sqrt{3}/2\simeq 0.866$ exactly reproduces the first two terms of the exact $A_\phi$ (\ref{ExactLoop}) for a single loop. The two mirrors are slightly different. Our fit gave $a=45.238\,\mathrm{mm}$, $\lambda = 0.9019$ and a loop separation of $8.251\,\mathrm{mm}$ between the two coils of the left mirror, and $\lambda=0.9027$ and a loop separation of $8.579\,\mathrm{mm}$ between the two coils for the right mirror, and a separation between the two mirrors of $274\,\mathrm{mm}$.  We found that these choices gave $\max(|\mathbf{B}_\mathrm{fit}-\mathbf{B}_\mathrm{exact}|)<0.02\,\mathrm{T}$ when the mirror field was $\sim 1\,\mathrm{T}$. This maximum error occurred on the wall of the trap directly underneath the mirrors; for $\sqrt{x^2+y^2}<15\,\mathrm{mm}$ the maximum error was $\sim 0.01\,\mathrm{T}$.

\subsection{Octupole field}

The vector potential for an infinite octupole is
\begin{equation}
\mathbf{A}_\infty = F s^4\cos (4\phi )\hat{z},
\end{equation}
where $s=\sqrt{x^2+y^2}$ and $F$ is a constant. A finite, symmetric octupole can be written as
\begin{equation}
A_z = (F_4(z)s^4 + F_6(z)s^6 + F_8(z)s^8+...)\cos (4\phi ),
\end{equation}
where the $F$'s are functions to be determined later. The condition
\begin{equation}
\nabla^2A_z = 0
\end{equation}
gives the relations
\begin{eqnarray}
F_6 &=& -\frac{F_4''}{20},\\
F_8 &=& -\frac{F_6''}{48}=\frac{F_4^{(iv)}}{960},\\
&\mathrm{etc.}&\nonumber
\end{eqnarray}

In order to satisfy $\mathbf{\nabla}\cdot\mathbf{A}=0$ there must be non-zero
components of $\mathbf{A}$ in the $s$ and $\phi$ directions:
\begin{eqnarray}
A_s &=& (G_5(z)s^5 + G_7(z)s^7+...)\cos (4\phi ),\\
A_\phi &=& (H_5(z)s^5+H_7(z)s^7+...)\sin (4\phi ).
\end{eqnarray}
The $G$'s and $H$'s are determined by the equations
\begin{eqnarray}
\mathbf{\nabla}\cdot \mathbf{A} &=& 0=\frac{1}{s}\frac{\partial}{\partial s}(sA_s)
+\frac{1}{s}\frac{\partial A_\phi}{\partial\phi} +\frac{\partial A_z}{\partial z},\\
\nabla^2A_x &=& 0,\\
\nabla^2A_y &=& 0.
\end{eqnarray}
The second two relations lead to $G_5=H_5$, $G_7=H_7$ etc. The first relation
leads to
\begin{eqnarray}
G_5(z) &=& -\frac{1}{10}F'_4(z),\\
G_7(z) &=&-\frac{1}{12}F_6'(z)= \frac{F_4'''(z)}{240},\\
&\mathrm{etc.}&\nonumber
\end{eqnarray}
Note that there is only one free function, $F_4(z)$; all of the other
functions are derivatives of this one. To get a fit to $F_4$ we need a
function that looks like a symmetric plateau. We chose to use the
complementary error function
\begin{equation}
F_4(z) = D [\mathrm{erfc}((z-z_f)/\Delta z) - \mathrm{erfc}((z+z_f)/\Delta z)],
\end{equation}
where $D$ is a constant, $\pm z_f$ are the approximate ends of the
octupole and $\Delta z$ is the distance in $z$ over which the octupole
drops to $\sim 0$.

In our fit, we found $z_f=129.46\,\mathrm{mm}$ and $\Delta z=16.449\,\mathrm{mm}$.  This form was able to get $\max(|\mathbf{B}_\mathrm{fit}-\mathbf{B}_\mathrm{exact}|)<0.02\,\mathrm{T}$ when the field at the wall was $\sim 1.5\,\mathrm{T}$. Because of the way in which the functions were chosen, the condition $\mathrm{\nabla}\cdot\mathbf{A}$ is always exactly satisfied if the orders of the expansion are kept the same in all three components of $\mathbf{A}$. The condition $\nabla^2\mathbf{A}=0$ is satisfied only to the extent that enough terms are retained in the expansion.  For our parameters, $\nabla^2\mathbf{A}$ is small in the region of interest inside the trap.

\section{Mirror-Trapped Antiproton Trajectories}
\label{pbarOrbits}
\begin{figure}[thb]
\centerline{\includegraphics{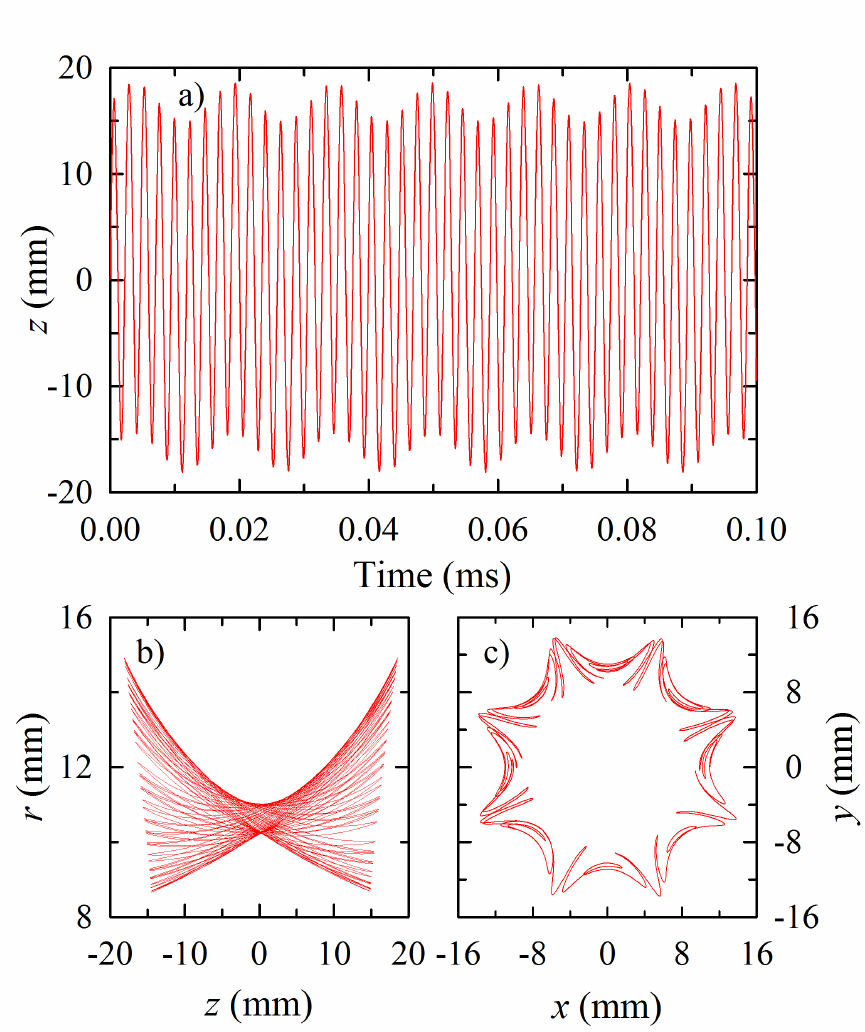}}
\caption{Typical trajectory of an electrostatically trapped antiproton, with $\Epar=\Eperp=10\,\mathrm{eV}$ and a starting radius of $11\,\mathrm{mm}$. The magnetic fields are held constant  at the values given in Section~\ref{Introduction}. The antiproton oscillates in a well formed by two end electrodes biased to $-140\,\mathrm{V}$, separated by a $80.2\,\mathrm{mm}$ grounded electrode. As in Figure~\ref{apparatus}, the trap's central axis points along $\zhat$, and the center of the trap, at $z=0$, is in the center of the grounded electrode.   a) Axial ($t$--$z$), b) radial ($z$--$r$), and c) transverse ($x$--$y$) projections of the motion.}
\label{pbar_electrostatic}
\end{figure}

Electrostatically trapped antiprotons follow regular trajectories similar to those shown in Figure~\ref{pbar_electrostatic}.  The antiprotons oscillate between the two ends of the electrostatic well, following field lines that typically extend between a radial minimum at one end of the well to a radial maximum at the other end of the well.  These radial maxima occur in magnetic cusps \cite{gomb:07,faja:08}, four to each side, caused by the octupole's radial fields.  Guiding center drifts cause the antiprotons to slowly rotate around the trap axis, so that the trajectories slowly alternate between cusps at each end.  The consequences of this motion, like the existence of a limit on the maximum allowed well length, have been explored in a series of papers \cite{faja:04,faja:05,gomb:07,andr:07,faja:08}.

\begin{figure}[thb]
\centerline{\includegraphics{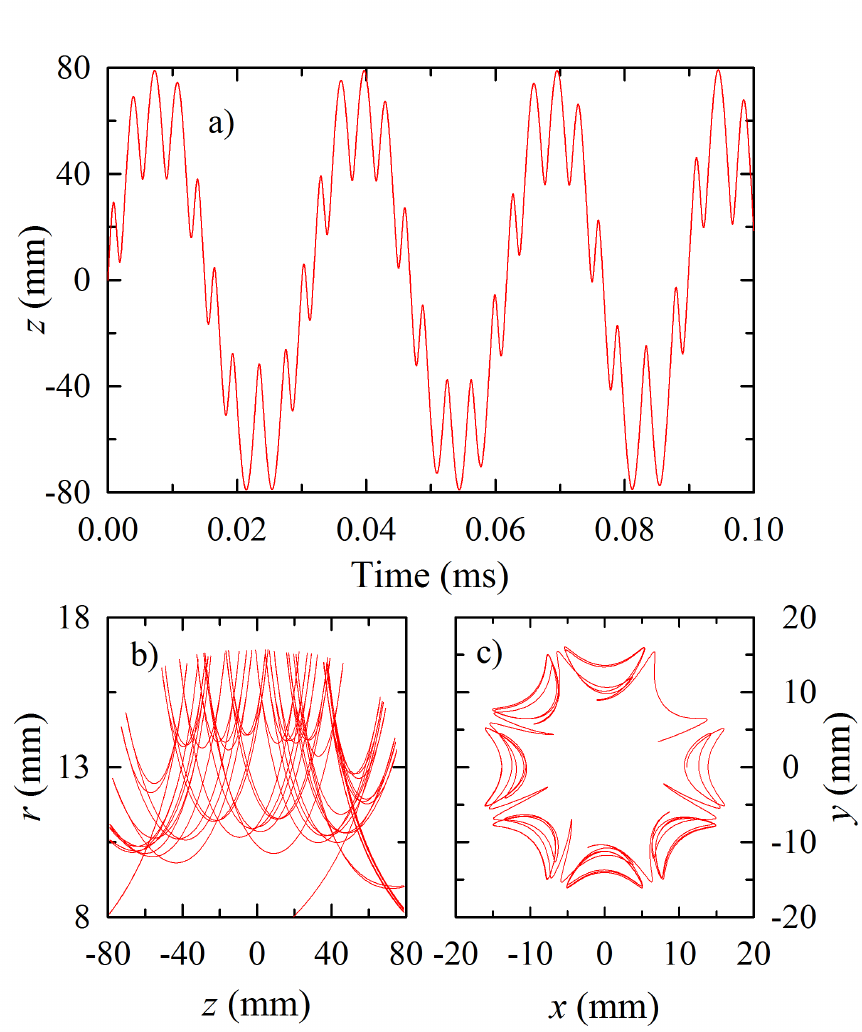}}
\caption{Typical trajectory of a mirror-trapped antiproton, with $\Epar=10\,\mathrm{eV}$, $\Eperp=60\,\mathrm{eV}$ and a starting radius of $11\,\mathrm{mm}$.  All electrodes are grounded.  a), b) and c) are described in Figure~\ref{pbar_electrostatic}.}
\label{pbar_mirror}
\end{figure}

\begin{figure}[thb]
\centerline{\includegraphics{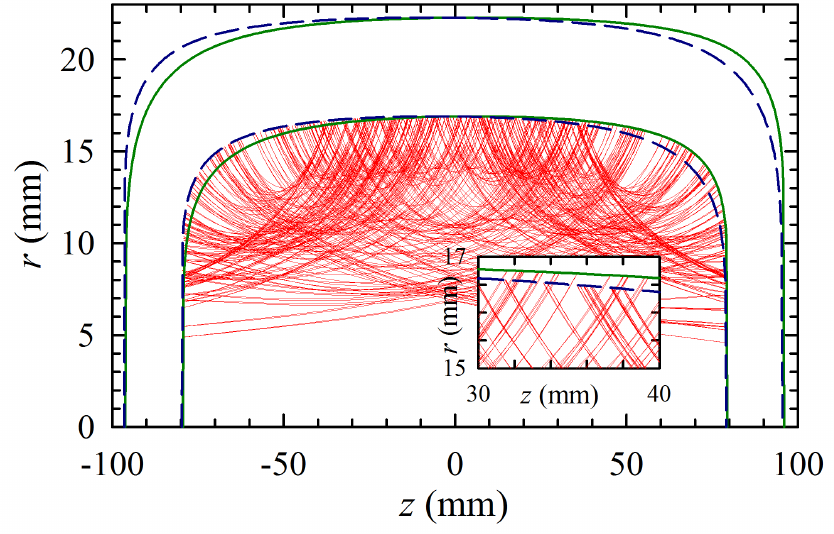}}
\caption{The same trajectory as in Figure~\ref{pbar_mirror}b, but plotted for $1\,\mathrm{ms}$ rather than $0.1\,\mathrm{ms}$.  The solid and dashed lines are lines of constant $|\mathbf{B}|$, plotted at the two angles, $22.5^{\circ}$ and $-22.5^{\circ}$, of the octupole cusps.  One set of lines is plotted at $|\mathbf{B}|=1.26\,\mathrm{T}$, the reflection field for the plotted trajectory.  The other set is plotted at the value of $|\mathbf{B}|$ such that the largest radial extent equals the wall radius $\Rw$ at $10\,\mathrm{ms}$ after the magnet shutdown, a typical time for an antiproton to hit the wall. The inset figure shows that the trajectories terminate on one angle or the other depending on their $z$ direction. }
\label{Bmag_reflection}
\end{figure}

Mirror-trapped antiprotons trace far more complicated trajectories, as shown in Figure~\ref{pbar_mirror}.  Typically, the $z$ motion follows a relatively slow macro-oscillation that extends over the full axial extent, and a faster micro-oscillation, over a more limited axial extent.  Each micro-oscillation typically travels between two large-radius, local octupole cusps, though sometimes the micro-oscillation ends at a low-radius point near one of the mirror coils.  Since the only mechanism for reversing the antiproton motion is a magnetic mirror reflection, the reversal necessarily occurs at a relatively large value of $|\mathbf{B}|$.  Indeed, the reflection always occurs at the {\em same} value of $|\mathbf{B}|$: at the field magnitude at which all of the antiproton's kinetic energy is completely tied up in its conserved magnetic moment $\mupbar$ (see Figure~\ref{Bmag_reflection} and (\ref{MirrorField})).

As can also be seen in Figure~\ref{Bmag_reflection}, the trajectories take the antiprotons closest to the trap wall in the center of the trap.  The electric field sloshing in the clearing cycles leaves the antiprotons with $\Epar$ on the order of $5$--$10\,\mathrm{eV}$.  Consequently, the antiprotons oscillate from one end of the trap to the other rapidly; for the trajectory in Figure~\ref{pbar_mirror}, the macro-oscillation bounce frequency is on the order of $20\,\mathrm{kHz}$.  After the magnet shutdown, antiprotons escape over a time of more than $10\,\mathrm{ms}$; thus, the antiprotons typically make hundreds of bounces during the shutdown process.    This allows the antiprotons to find the ``hole'' in the trap center, and causes the antiproton annihilations to be concentrated there in the No Bias case (see Figure~\ref{BenchmarkPlot}).  When a Bias is applied, the center of the pseudopotential moves to the side (Figure~\ref{QWP_Potentials}), and the annihilation center follows.

\section{Minimum-B Trapped Antihydrogen Trajectories}
\label{HbarOrbits}
\begin{figure}[thb]
\centerline{\includegraphics{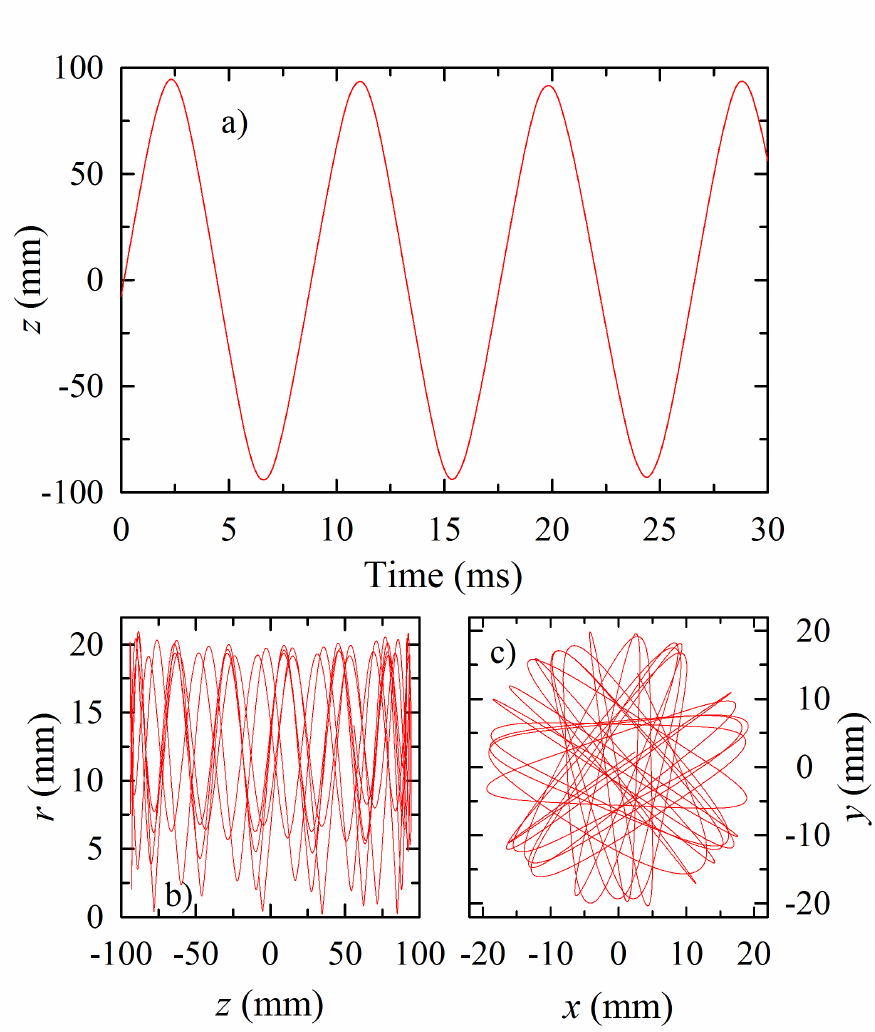}}
\caption{Typical trajectory of a minimum-B trapped antihydrogen atom.  The antiatom started with a kinetic energy of $0.5\,\mathrm{K}$.  a), b) and c) are described in Figure~\ref{pbar_electrostatic}.}
\label{Hbar_orbit}
\end{figure}

A typical minimum-B trapped antihydrogen atom trajectory is graphed in Figure~\ref{Hbar_orbit}.  In the transverse plane, the antiatom oscillates radially, with a varying rotational velocity; a Fourier transform (not shown) of the $x$ or $y$ motion yields a broad range of frequencies.  This is expected as an $r^3$ potential in the transverse plane, to which the potential in our trap approaches, is known to yield chaotic motion \cite{khas:07}.

\begin{figure}[thb]
\centerline{\includegraphics{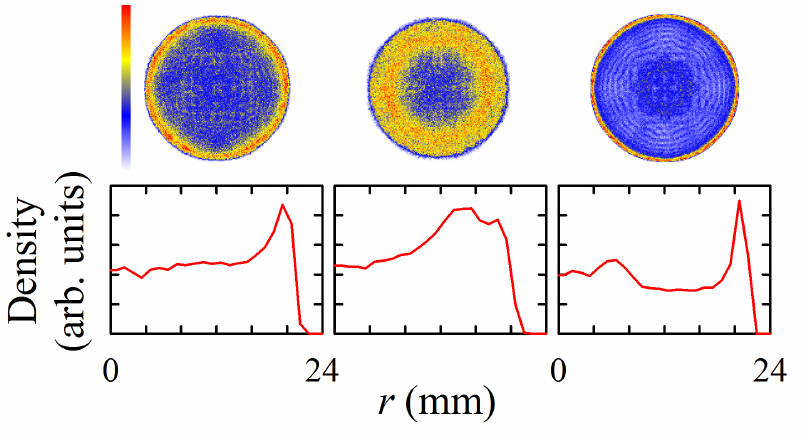}}
\caption{Transverse projections of three typical antihydrogen atom trajectories, each for antiatoms with $0.5\,\mathrm{K}$ energy, but with differing, randomly picked, initial directions. The antiatoms were propagated for $100\,\mathrm{s}$. Each projection was scaled to the same maximum on the linear colourmap.  Below each projection is the corresponding density profile.}
\label{Hbar_Radial_Profile}
\end{figure}

Typically, the antiatom trajectories cover the transverse plane reasonably uniformly, with little azimuthal structure, but are peaked at the outer radial edge where the antihydrogen atoms reflect (see Figure~\ref{Hbar_Radial_Profile}).  The ultimate goal of these experiments is, of course, to use spectroscopy to search for differences between antihydrogen and normal hydrogen. The plots in Figure~\ref{Hbar_Radial_Profile} suggest that the trajectories do not sharply constrain the waist of a probe laser or microwave beam.

\begin{figure}[thb]
\centerline{\includegraphics{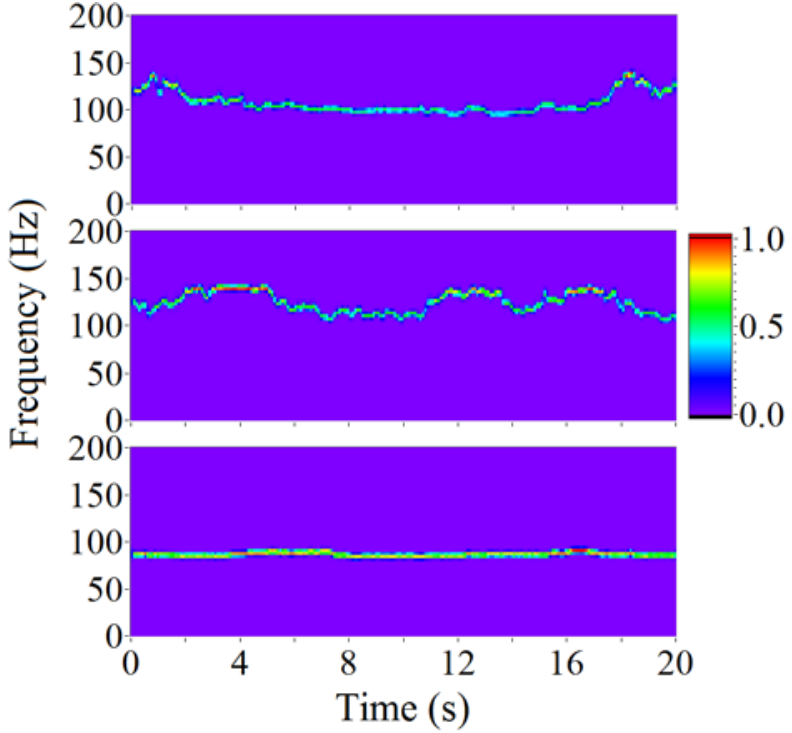}}
\caption{Sliding Fourier transform of the z-motion of the antihydrogen trajectories in Figure~\ref{Hbar_Radial_Profile}. The colour scale is linear.}
\label{z-Spectrum}
\end{figure}

The axial motion is quasi-harmonic with a well defined oscillation frequency that typically remains constant for many oscillations.  Occasionally, as shown in Figure~\ref{z-Spectrum}, the frequency jumps due to interactions with the transverse motion.  Since the $z$-oscillation frequency is on the order of $100\,\mathrm{Hz}$, the antiatoms bounce only a few times during the magnet shutdown.  Unlike mirror-trapped antiprotons, the antiatoms do not have time to find the low $|\mathbf{B}|$ hole in the $z$-center of the trap, and, consequently, they annihilate over a broad region in $z$.

\section{Ionization of fast antihydrogen}
\label{Bstrip}

To study the ionization of fast moving antihydrogen atoms, such antiatoms were propagated in a constant axial magnetic field $B=1\,\mathrm{T}$, and a radial electric field ${\bf E}=n_e q\boldsymbol \rho/2\epsilon_0=E_0{\boldsymbol \rho}$
arising from the space charge of the positron plasma. Here, $n_e=5.5\times10^7\,\mathrm{cm}^{-3}$ is the plasma density, and ${\boldsymbol \rho}=(x,y,0)$. The equations of motion in terms of the center of mass coordinates $\mathbf{R}_\mathrm{CM}$, $\mathbf{V}_\mathrm{CM}$, and the relative coordinates $\mathbf{r}$, $\mathbf{v}$, are
\begin{eqnarray}
  M\dot\mathbf{V}_\mathrm{CM} &=& q \mathbf{v} \times \mathbf{B}+ q E_0{\boldsymbol \rho} \label{eq:E1}\\
  \mu\dot\mathbf{v} &=& q (\mathbf{V}_\mathrm{CM}+\lambda\mathbf{v}) \times \mathbf{B}+ q E_0({\boldsymbol \rho}_\mathrm{CM}+\lambda {\boldsymbol \rho})+{\bf F}_c \label{eq:E2}\\
&=& q \lambda\mathbf{v} \times \mathbf{B}+ q {\bf E}_{\rm eff}+{\bf F}_c, \nonumber
\end{eqnarray}
where $M$ is the total mass of the atom, $\mu$ the reduced mass,
$\lambda=(m_p-m_e)/M$ and ${\bf F}_c$ the Coulomb force. The effective
electric field ${\bf E}_{\rm eff}= \mathbf{V}_\mathrm{CM} \times \mathbf{B} +
qE_0({\boldsymbol\rho}_\mathrm{CM}+\lambda{\boldsymbol \rho})$ is the sum of the regular
electric field and a term
proportional to the center-of-mass velocity of the atom.

The coupled equations (\ref{eq:E1}) and (\ref{eq:E2}) were solved using an adaptive step size Runge-Kutta algorithm. The antihydrogen atoms were initialized at a trap radius of $0.5\,\mathrm{mm}$ and some given initial binding energy $\Ebind$ and initial kinetic energy  $\Ekin$ in the transverse plane. Binding energy is here defined as in the field-free situation, i.e.\ as the sum of the kinetic energy of the positron and the Coulomb potential. For each parameter set $\{\Ebind,\Ekin\}$, one thousand trajectories were calculated. Each trajectory was followed for a maximum of $2\,\mu\mathrm{s}$ or until the atom was ionized. The fraction of trajectories leading to ionization, as well as the time until ionization, were recorded.

The magnetic field creates an effective harmonic confinement for the positron in the transverse plane. Hence, strictly speaking, one cannot have field-ionization (in the sense that $r\rightarrow\infty$), unless there is also some axial electric field present, which was not the case in these simulations. However, a strong radial electric field will induce a positron-antiproton separation much larger than the atomic size in the field-free situation. Such a positron will be bound only by a negligible binding energy, and the antiatom will almost instantly be destroyed by either collisions with another positron (inside the positron plasma) or by a weak axial electric field (just outside the plasma). We regard any antihydrogen atom bound by less than $2\,\mathrm{K}$, which is much less than the plasma temperature, as ionized.

\begin{figure}[tb]
\centerline{\includegraphics{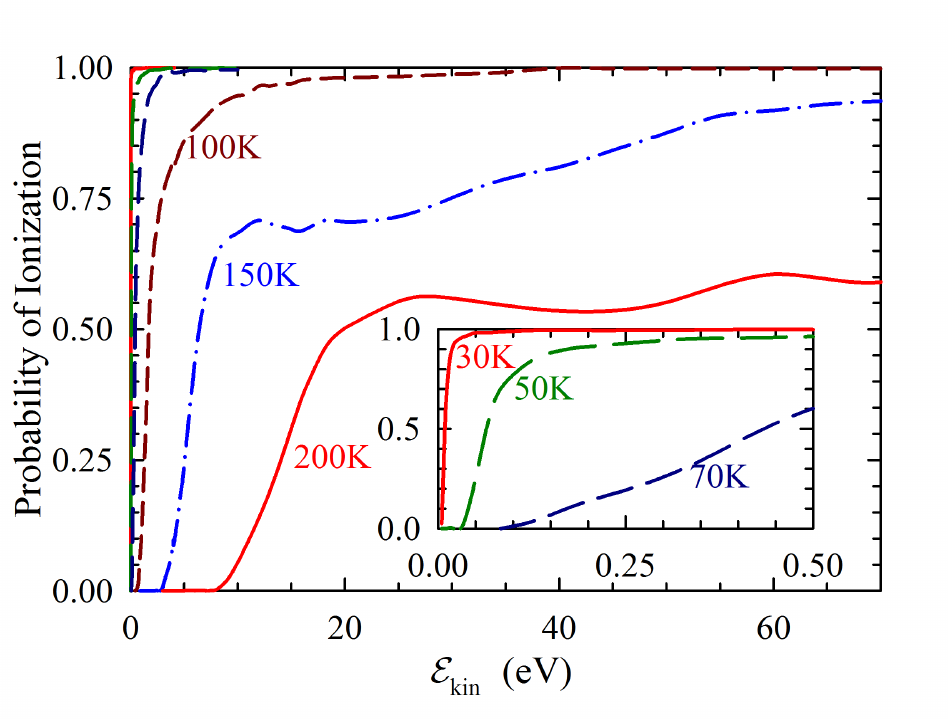}}
  \caption{Probability of ionization within $2\,\mu\mathrm{s}$ as a function of kinetic energy for antihydrogen atoms with the listed binding energies $\Ebind$.}
  \label{fig:E1}
\end{figure}

The fraction of antihydrogen trajectories leading to ionization is shown in Figure~\ref{fig:E1} for various initial binding energies and center-of-mass velocities. An antihydrogen atom is stable against ionization by an axial electric field $E_z$ for binding energies $\Ebind>2\sqrt{(q^2/4\pi\epsilon_0)q E_z}$. Typical electric fields in the trap are of order $10\,\mathrm{V}/\mathrm{cm}$, corresponding to stability for $\Ebind\gtrsim 30\,\mathrm{K}$. Any antihydrogen atom with a binding energy less than $30\,\mathrm{K}$ will be ionized by the effective electric field with more than 99\% efficiency at even moderate kinetic energies of $0.1\,\mathrm{eV}$. However, very close to the electrode boundaries, the electric fields can be much larger, corresponding to stability only for $\Ebind> 150\,\mathrm{K}$. Our simulations show that such deeply bound atoms will require much larger kinetic energies to ionize in the lower-field region in the center of the trap (see Table~\ref{tab:E1}).

\begin{table}[h]
  \caption{\label{tab:E1}Minimum kinetic energy $\Ekin$ of an antihydrogen
    atom required
    for ionization within 2$\mu$s with 90\% probability (column 2) and
    99\% probability (column 3) for different binding energies $\Ebind$.}
\begin{indented}
  \item[]\begin{tabular}{c@{\extracolsep{3mm}}cc}
  \br
    $\Ebind\,(\mathrm{K})$ & \multicolumn{2}{c}{$\Ekin\,(\mathrm{eV})$} \\

       &   90\% & 99\% \\
    \mr
    30 &  0.02 & 0.1 \\
    40 &  0.06 & 0.7  \\
    50 &  0.2 & 1.5 \\
    60 &  0.7 & 2.3  \\
    70 &  1.4 & 4.1 \\
    80 &  2.3 & 10 \\
    90 &  4   & 25\\
    100 & 7   & 35\\
    150 & 55  & 150\\
    \br
  \end{tabular}
   \end{indented}
\end{table}

\section{Energy Reconstruction}
\label{EnergyReconstruction}
The trapped antihydrogen energy distribution function $f(\Energy)$ can be crudely reconstructed from the time history of the annihilations after the magnet shutdown.  As shown in Figure~\ref{EnergyDist}b, antiatoms of a given energy $\Energy$ annihilate over a broad distribution of times.  The overall probability distribution function for the antiatoms to annihilate at time $t$ can be found by integrating the probability $P(t|\Energy)$ of annihilation at time $t$ of antiatoms with specific energy $\Energy$ over the antiatom energy distribution function:
\begin{equation}
\label{Recon_forward}
f(t)=\int_0^\infty \mathrm{d}\Energy\,P(t|\Energy) f(\Energy).
\end{equation}
This equation can be exploited by guessing a distribution function $f(\Energy)$, calculating $P(t|\Energy)$ with simulations, and comparing it to a histogram of the observed data. This ``forward'' method was explored in \cite{andr:11a} and in Figure~\ref{EnergyDist}c.

Alternately, we can write
\begin{equation}
\label{Recon_inverse}
f(\Energy)=\int_0^\infty \mathrm{d}t\,P(\Energy|t) f(t).
\end{equation}
In this appendix we explore the consequences of employing this ``inverse'' equation.  We perform the integral in (\ref{Recon_inverse}) as follows: (a) For each annihilation event, construct a narrow band around the annihilation time in the antihydrogen simulation results.  A typical such band is shown in gray in Figure~\ref{EnergyDist}b.  (b) From this band, randomly select a fixed number of the simulated annihilation events. (We selected 20 such samples in the reconstructions in this paper.) This effectively finds and samples $P(\Energy|t)$.  (c)  Aggregate all the energies from the randomly selected samples for each observed event, effectively integrating over $t$ as properly weighted by $f(t)$.   (d) Construct the histogram of these aggregated samples; this is the reconstructed energy distribution.

\begin{figure*}[thb]
\centerline{\includegraphics{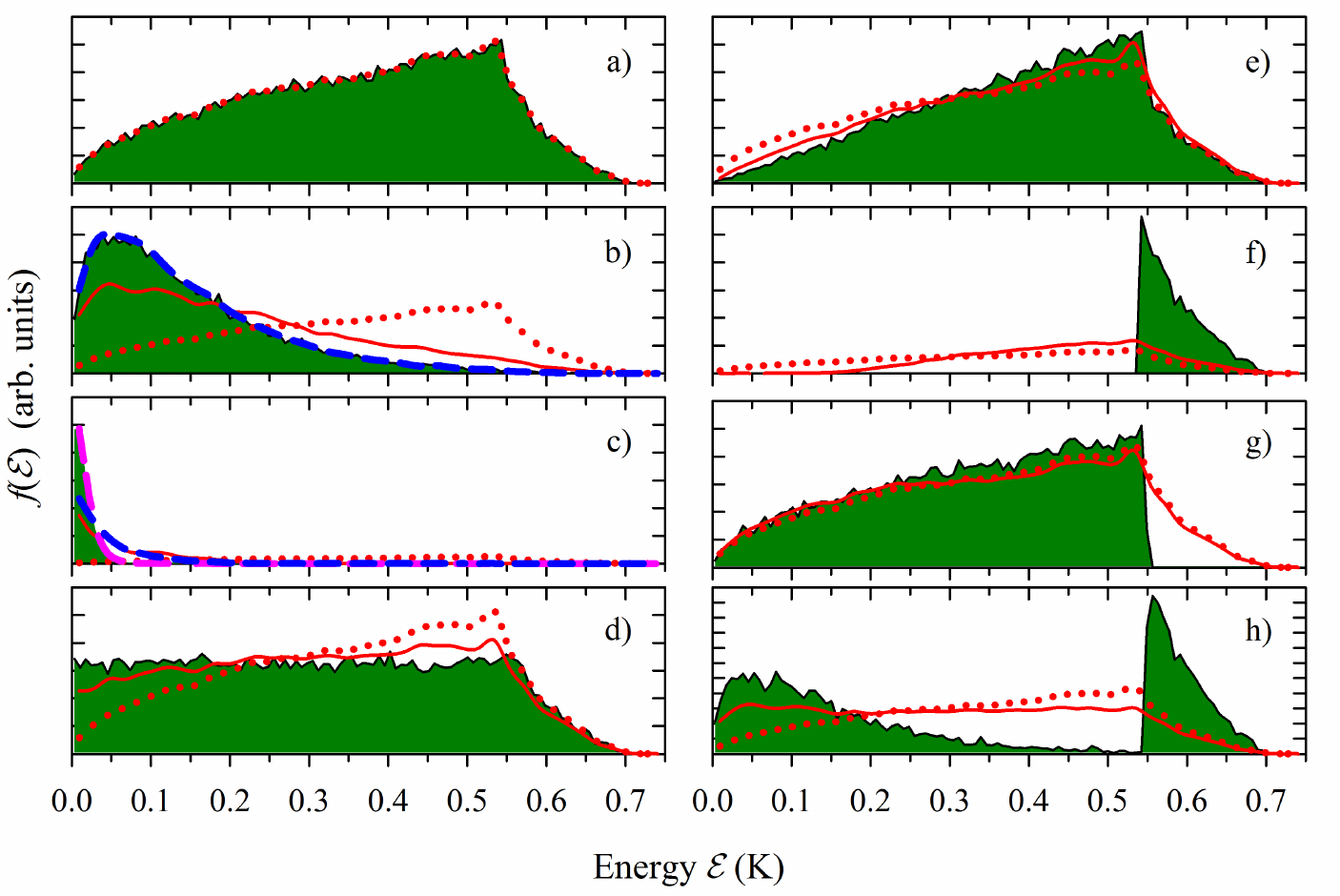}}
\caption{Monte-Carlo study of the trapped antihydrogen distribution function reconstruction algorithm. In all eight subgraphs, the green, solid region is a histogram of a postulated distribution function $f(\Energy)$ whose reconstruction (red, blue and pink lines) is being attempted.  The histograms are generated from: a) the distribution of surviving (i.e.\ trapped) antihydrogen atoms as predicted by the antihydrogen simulation from a starting population of atoms at $54\,\mathrm{K}$.  This is the distribution principally studied in this paper.  In this and in all subsequent cases, the histograms are not smooth because they are generated from a finite number of samples from the starting population; b) the distribution of surviving antihydrogen atoms starting from a population of atoms at $0.1\,\mathrm{K}$; c) the distribution of surviving antihydrogen atoms starting from a population of atoms at $0.01\,\mathrm{K}$;  d) a distribution similar to that in a), but with the distribution artificially forced to be flat out to the trapping energy, and then rolled off with the same quasi-bound distribution as found in a);  e) a distribution that is similar to d), but which is artificially forced to increase linearly out to the trapping energy; f) the quasi-bound antiatoms in a) only;  g) the non-quasi-bound antiatoms in a) only. h) a double humped distribution.   In a), the red dotted line is the average reconstructed distribution function found using an inversion based on the $54\,\mathrm{K}$ simulation study, as described in \ref{EnergyReconstruction}.  For comparative purposes, this red dotted line is replicated in all the subgraphs. In all the subgraphs but the first, the red solid line is the reconstruction of the postulated distribution function in the particular subgraph, also found with an inversion based on the $54\,\mathrm{K}$ simulation study.  In b) and c), the blue dashed lines are reconstructions found with an inversion based on a $100\,\mathrm{mK}$ simulation study.  In c), the pink dashed line is the reconstruction found with an inversion based on a $10\,\mathrm{mK}$ simulation study.  In each of the plots, the reconstructions are averaged over two thousand 309 point Monte-Carlo generated event sets.   (Figure~\ref{EnergyDist}a shows the typical reconstruction and error with just one 309 point event set---our actual data.) From this survey, it is clear that the mean energy of the distribution, as well as some coarse features of the distribution, are recovered, but sharp features are lost.}
\label{ReconstructedDistributions}
\end{figure*}

The eight subgraphs of Figure~\ref{ReconstructedDistributions} show a study of the reconstruction process for eight trial distributions.  For each trial distribution, we analyzed two thousand sets of 309 Monte-Carlo generated annihilation events, each event obeying the trial distribution particular to the figure subgraph.  The average over all of the resulting reconstructions for each subgraph is then plotted.  One can see that the reconstruction is coarse. Figures~\ref{ReconstructedDistributions}a--e show that the mean energy of the distribution is recovered approximately, as well as some features of the higher moments of the distribution, but, for more pathological distributions, Figures~\ref{ReconstructedDistributions}f--h show that the ability to recover these higher moments is limited.

The reconstruction errors stem from two causes: (1) The band of energies at each time is broad (see Figure~\ref{EnergyDist}b).  This results in sharp features being smeared;  this problem is particularly relevant in Figures~\ref{ReconstructedDistributions}f--h.  (2) The reconstruction has a difficult-to-quantify memory of the original distribution used in the simulations underlying the reconstruction; this problem is particularly acute in Figures~\ref{ReconstructedDistributions}b-c.  The reconstruction can be improved by iteration; an initial reconstruction, done employing the original simulation results, can be used to determine the approximate temperature of the experimental data, and then the reconstruction rerun using a simulation with a more appropriate temperature.

\clearpage



\end{document}